\documentclass[sort&compress]{elsarticle}

\usepackage{lineno,hyperref}
\usepackage{physics}

\makeatletter
\def\set@curr@file#1{%
	\begingroup
	\escapechar\m@ne
	\xdef\@curr@file{\expandafter\string\csname #1\endcsname}%
	\endgroup
}
\def\quote@name#1{"\quote@@name#1\@gobble""}
\def\quote@@name#1"{#1\quote@@name}
\def\unquote@name#1{\quote@@name#1\@gobble"}
\makeatother
\usepackage{graphicx}
\usepackage{floatrow}
\usepackage{bm}
\graphicspath{ {./Figures/} }
\journal{International Journal of Mechanical Sciences}

\usepackage{amssymb}
\usepackage{amsmath}
\usepackage{empheq}

\newcommand{\euler}{\mathrm{e}}

\newcommand{\laguerreL}[1]{L_{#1}}

\newcommand{\chebyshevU}[1]{U_{#1}}

\newcommand{\christovCC}[1]{CC_{#1}}

\makeatletter
\newcommand{\pushright}[1]{\ifmeasuring@#1\else\omit\hfill$\displaystyle#1$\fi\ignorespaces}
\newcommand{\pushleft}[1]{\ifmeasuring@#1\else\omit$\displaystyle#1$\hfill\fi\ignorespaces}
\makeatother
\DeclareMathOperator{\eulerGamma}{\Gamma}

\newcommand{\specialcell}[2][c]{%
	\begin{tabular}[#1]{@{}c@{}}#2\end{tabular}}
\begin{document}

\begin{frontmatter}
	
\title{Determination of Young's modulus of samples of arbitrary thickness from force distance curves: numerical investigations and simple approximate formulae}
\author[ph1,ph2]{Pawe\l{} Hermanowicz}
\ead{pawel.hermanowicz@uj.edu.pl}

\address[ph1]{Laboratory of Photobiology, Malopolska Centre of Biotechnology, Jagiellonian University, Gronostajowa 7A, 30-387 Krakow, Poland}

\address[ph2]{Department of Plant Biotechnology, Faculty of Biochemistry, Biophysics and Biotechnology, Jagiellonian University, Gronostajowa 7, 30-387 Krakow, Poland}

\begin{abstract}
We present simple expressions for load required to indent a layer of arbitrary thickness with a conical, paraboloidal or cylindrical punch. A rigid substrate underneath the sample leads to an increase of load required for indentation. This effect has to be corrected for to prevent overestimation of Young's modulus from indentation measurements, such as force - distance curves recorded with the Atomic Force Microscope (AFM). The problems of the frictionless contact of an axisymmetric punch and an isotropic, linear-elastic layer are reducible to Fredholm integral equations of the second kind. We solved them numerically and used the Remez algorithm to obtain piecewise polynomial approximations of the load – indentation relation for samples that are either in frictionless contact with the rigid substrate or bonded to it. Their relative error due to approximation is negligible and uniformly spread. Combining the numerical approximations with asymptotic solutions for very thin layers, we obtained equations appropriate for samples of arbitrary thickness. They were implemented in a new version of AtomicJ, our free, open source application for analysis of AFM recordings.

\end{abstract}


\end{frontmatter}

%%\linenumbers

\section{Introduction}
Mechanical properties of materials can be investigated through measurements of load required to indent a sample with a rigid punch. Such experiments are often performed with a nanoindenter or the Atomic Force Microscope (AFM). In AFM, a tip mounted at the free end of a cantilever acts as a rigid punch. The recorded relation between the position $z$ of the fixed base of the cantilever and the force $P$ acting on the tip pressed into the sample is known as a force - distance curve. The presence of a rigid substrate underneath the sample leads to an increase of load required for indentation. If this effect is not taken into account during data analysis, the obtained values of Young's modulus are overestimated \cite{Domke1998},\cite{Akhremitchev1999}. This makes it difficult to separate the effects of topography from the true variability of the mechanical properties of the sample, e.g. in the case of animal cells forming thin cytoplasmic protrusions \citep{Rotsch1999}. 

Indentation testing is often a method of choice for studies of elastic properties of  biological samples and biomedical materials due to their heterogeneity and the necessity to keep the sample in its intact form. Studies of biological tissues, such as articular cartilage, also served as motivation for theoretical progress in studies of the effect an underlying substrate on the apparent sample stiffness in indentation tests \citep{Hayes1972, Argatov2013, Argatov2013b}. Practical applications of indentation testing include examination of changes in cell stiffness associated with progression of diseases such as cancer \citep{Lekka1999, Li2008}, infection \citep{Eaton2012} and asthma \citep{Sarna2015}. Taking into account the effect of the substrate may aid in the interpretation of experiments in which stiffness of different cell lines or types are compared, reducing confounding due to differences in the cell geometry or adherence to the substrate. Accurate determination of Young's modulus of thin layers from indentation experiments is also important in tissue engineering applications. For example, adhesion and area of cells growing on thin films deposited on other substrates depend on Young's modulus of these films \citep{Schneide2006, Thompson2005}.
  
To extract mechanical properties of the sample from indentation measurements, it is necessary to assume a theoretical model of its contact with the punch. The theory of frictionless contact between a punch and a layer supported by a substrate is among the most studied topics in contact mechanics. The problem of an axisymmetric punch pressed into a non-bonded, isotropic layer, resting on a rigid half-space, was studied by Lebedev and Ufliand \cite{Lebedev1958}, who reduced it to a Fredholm integral equation of the second kind. Similar studies have been carried out for a layer bonded to the rigid substrate \citep{Pupyrev1960} and a layer bonded to a compliant, isotropic half-space \cite{Dhaliwal1970b,Dhaliwal1970,Rau1972,Yu1990, Korsunsky2009}. The more general problem of a stratified layer, supported by a half-space, has also received much attention \citep{Chen1972a, Chen1972b, Kuo1992, Maltis2005, Constantinescu2013}. Indentation of non-linear elastic layers, supported by rigid substrates, has been also studied, using the Finite Element Method \cite{Finan2014, Fessel2017}.

The magnitude of the effect of the rigid substrate depends on the ratio $\tau$ of the contact radius $a$ (the radius of the area of direct contact between the punch and the sample, see fig. \ref{fig:deformationThinSample}) and the sample thickness $h$. Methods for derivation of asymptotic formulae for load $P$ and indentation $\delta$ have been presented in \cite{Vorovich1959, England1962, Vorovich1974} (for $\tau \ll 1$) and in \cite{Aleksandrov1969} (for $\tau \gg 1$). These expressions have a parametric form - both load and indentation depth are expressed as functions of a parameter which is not directly measured, so that numerical calculations are necessary to find the load - indentation relation. Formulae for load expressed as an explicit function of $\delta$ are more convenient. Approximations of $P(\delta)$, based on the assumption of small $\tau$, have been derived in \cite{Dimitriadis2002, Argatov2011, Garcia2018} for a paraboloidal punch, and in \cite{Argatov2011,Gavara2012,Managuli2017,Managuli2018} for a conical punch. Analogous approximations of $P(\delta)$ appropriate for large values of $\tau$ have been proposed in \cite{Yang2003} for conical and paraboloidal indentation.

During development of AtomicJ \cite{Hermanowicz2014}, our open source application for analysis of AFM recordings, we became aware of the need to develop polynomial approximations of load - indentation functions for samples of arbitrary thickness, with a small and uniformly spread error. Force - distance curves are used to assess elastic properties of diverse samples, whose thickness varies from a few nanometres in the case of lipid layers to millimetres in the case of tissue samples. Thus, approximations appropriate for implementation in a publicly available software should remain accurate for a wide range of sample thickness, preferably for all non-negative values of the parameter $\tau$. 

In this paper, we present simple, piecewise polynomial approximations of $P(\delta)$ for frictionless contact of a conical, paraboloidal or cylindrical punch with an isotropic, linear elastic layer resting on a rigid substrate. Their error is low and uniform. The approximates have been implemented in AtomicJ. We also compare the numerical solutions with formulae known from the literature, including those frequently used to analyse force - distance curves. In addition, we examine the effect of the  rounded apex of conical punches on load required to indent a layer and present approximations for load valid when $\tau \ll 1$.

\section{Contact problem for an elastic layer}

\begin{figure}[!t]
	\floatbox[{\capbeside\thisfloatsetup{capbesideposition={left,bottom},capbesidewidth=5.3cm}}]{figure}[\FBwidth]
	{\caption{Indentation of a layer of thickness $h$, resting on a rigid substrate, with an axisymmetric punch. The depth of indentation is denoted as $\delta$, contact radius as $a$ and load as $P$. The $z$ axis is directed towards the substrate.}\label{fig:deformationThinSample}}
	{\includegraphics{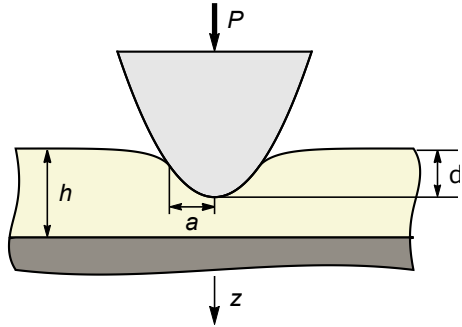}}
\end{figure}

We will consider a frictionless, normal contact between a linear-elastic layer of thickness $h$ and a rigid, axisymmetric punch (fig. \ref{fig:deformationThinSample}). The punch profile is described by a smooth function $f(\varrho)$, satisfying $f(0) = 0$. The upper surface ($z = 0$) of the layer is loaded by the punch. The lower surface ($z = h$) is supported by a rigid substrate of infinite thickness. Due to the axial symmetry, the region of direct contact between the layer and the punch is a disk of radius $a$. Circumferential displacements vanish everywhere. The boundary conditions on the upper ($z = 0$) surface of the layer are
\begin{align}
&u_z(\varrho,0)=\delta-f(\varrho)&0\leqslant\varrho\leqslant 1
\label{eq:axiSymContactBoundaryA}\\&\sigma_z(\varrho,0)=0&\varrho>1\label{eq:axiSymContactBoundaryB}\\&\tau_{rz}(\varrho,0)=0\label{eq:axiSymContactBoundaryC}
\end{align}
where $\varrho = \frac{r}{a}$ is the normalized radial coordinate, $u_z(\varrho,z)$ is normal displacement, $\sigma_z(\varrho,z)$ is normal stress, $\tau_{rz}(\varrho,z)$ is shear stress, $\delta$ is the depth of indentation. The conditions at the lower surface depend on the type of the contact with the substrate. If the contact with the substrate is frictionless and the displacement of points on the lower sample surface is restricted only in the normal direction, then
\begin{align}
&u_z(\varrho,h)=0
\label{eq:axiSymContactBoundaryFiniteA}\\&\tau_{r z}(\varrho,h)=0\label{eq:axiSymContactBoundaryFiniteB}
\end{align}
If the sample is bonded to the substrate, the points in contact with the substrate cannot move in the radial or normal direction
\begin{align}
&u_z(\varrho,h)=0
\label{eq:axiSymContactBoundaryFiniteBondedA}\\&u_r(\varrho,h)=0\label{eq:axiSymContactBoundaryFiniteBondedB}
\end{align} 
The contact problems specified by the above conditions resemble the classical Hertz problem for a punch of a circular planform, except for the finite thickness of the sample. The limitations of the Hertz-type contact models and the alternative approaches to contact modelling are discussed in \cite{Borodich2014}. The problem of a non-bonded elastic layer, resting on a rigid substrate (conditions \eqref{eq:axiSymContactBoundaryA} -- \eqref{eq:axiSymContactBoundaryFiniteB}), was studied by Lebedev and Ufliand \cite{Lebedev1958}. The problem of a bonded layer, described by \eqref{eq:axiSymContactBoundaryA} -- \eqref{eq:axiSymContactBoundaryC} and \eqref{eq:axiSymContactBoundaryFiniteBondedA} -- \eqref{eq:axiSymContactBoundaryFiniteBondedB}, was studied by Pupyrev and Ufliand \cite{Pupyrev1960}. Both problems were reduced to a Fredholm integral equation of the second kind
\begin{equation}
\label{eq:finiteSampleDerZA}
\chi(x;\tau) = \chi^\infty(x) +	\frac{1}{\pi}\int_{0}^{1}\chi(t;\tau)K(x,t;\tau)\dd{t} 
\end{equation}
We will refer to \eqref{eq:finiteSampleDerZA} as the Lebedev-Ufliand equation. Its solution $\chi(x)$ can be used to calculate parameters describing the tip-sample contact, including load and contact radius. Indentation depth and the punch profile enter the equation through its free term $\chi^\infty(x)$
\begin{equation}
\label{eq:chi_Infinitely_Thick}
	\chi^\infty(x) = \frac{2}{\pi}\left[\delta- x\int_{0}^{x}\frac{\dv{f}{z} (z)}{\sqrt{x^2-z^2}}\dd{z}\right]
\end{equation}
The kernel $K$ of \eqref{eq:finiteSampleDerZA} can be expressed as
\begin{equation}
\label{eq:finiteSampleDerKernelA}
K(x,t;\tau)=\Omega(t+x;\tau)+\Omega(t-x;\tau)
\end{equation}
where $\Omega(t+x;\tau)$ is a cosine transform of the weight function $\omega(p)$
\begin{equation}
\label{eq:finiteSampleDerOmega}
\Omega(y;\tau)=\int_{0}^{\infty}\omega(p;\tau)\cos(p y)\dd{p}
\end{equation}
The type of contact between the layer and the substrate determines the form of $\omega(p)$.  For a non-bonded layer,  $\omega(p)$ depends on one parameter  $\tau = \frac{a}{h}$. It can be expressed as \cite{Lebedev1958}
\begin{equation}
\label{eq:finiteSampleDerJ3}
\omega(p;\tau) =1-\frac{2\sinh[2](\frac{ p}{\tau})}{\frac{2 p}{\tau}+\sinh(\frac{2 p}{\tau})}
\end{equation}
For a bonded layer, $\omega(p)$ depends on two parameters, $\tau$ and Poisson's ratio $\nu$ \cite{Pupyrev1960}
\begin{equation}
\omega_b(p;\tau,\nu)=\frac{(3-4\nu)^2+(1+2\frac{p}{\tau})^2+2(3-4\nu)\euler^{-2\frac{p}{\tau}}}{(3-4\nu)\euler^{2\frac{p}{\tau}}+(3-4\nu)^2+(1+4\frac{p^2}{\tau^2})+(3-4\nu)\euler^{-2\frac{p}{\tau}}}\label{eq:finiteBondedSampleDerG}
\end{equation}
$\Omega(y;\tau)$ approaches zero when $\tau \xrightarrow[]{}\infty$. In this limiting case, the solution $\chi$ becomes equal to $\chi^\infty$.

The solution $\chi$ of \eqref{eq:finiteSampleDerZA} can be used to calculate normal stress at the upper surface $\sigma_z(\varrho,0)$, according to the equation
\begin{align}
\label{eq:finiteSampleDerStressZE}
\sigma_z(\varrho,0)=-\frac{E}{2a(1-\nu^2)}\left(\frac{\chi(1)}{\sqrt{1-\varrho^2}}- \int_{\varrho}^{1}\frac{\dv{\chi(t)}{t}}{\sqrt{t^2-\varrho^2}}\dd{t}\right)&& \varrho \leqslant 1 
\end{align}
The load $P$ can be calculated by integration of $\sigma_z(\varrho,0)$ within the area of contact between the punch and the layer. The final expression is
\begin{equation}
\label{eq:axSymInfinitSolJ8}
P=\frac{\pi a E}{(1-\nu^2)}\int_{0}^{1}\chi(x)\dd{x}
\end{equation}
where $E$ is the Young's modulus of the layer.

The boundary conditions for the problem of a punch indenting a layer do not determine the contact radius $a$, i.e. the radius of the area of direct contact between the punch and the layer. If forces of adhesion are absent, an additional condition of finite stress at the edge of the contact area must be introduced. In accordance with \eqref{eq:finiteSampleDerStressZE}, it leads to the criterion \cite{Lebedev1958}
\begin{equation}
\label{eq:noAdhesionTauEquilibriumCriterion}
	\chi(1;\tau) = 0
\end{equation}
More general criteria for the contact radius are provided by the Johnson-Kendall-Roberts (JKR) model of adhesive contact, which can be extended in a natural way to the contact between a punch and a thin layer \citep{Choi2012,Zhu2017b}. 

\begin{figure}[!t]
	\floatbox[{\capbeside\thisfloatsetup{capbesideposition={left,bottom},capbesidewidth=5.5cm}}]{figure}[\FBwidth]
	{\caption{The profiles of axisymmetric punches considered in this work. A cone of  half-angle $\theta$ (a), a paraboloid of radius $R$ (b) and a flat-ended cylinder of radius $a$ (c) are special cases of a power-law-shaped punch. A blunt conical punch (d) can be described as a frustum of a cone of half angle $\theta$, merged with a paraboloidal apex of radius $R$. The radius of the transverse cross-section of the blunt cone at the joint between the apex and the conical body is denoted by $b$.}\label{fig:figFourPunchShapes}}
	{\includegraphics[width=6cm]{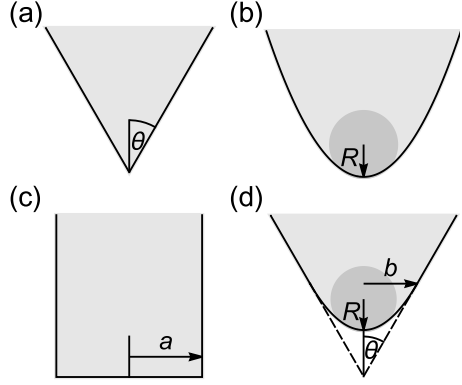}}
\end{figure}

\paragraph{Punch profiles}In this work, we will consider power-law-shaped punches, with profiles described by
\begin{equation}
\label{eq:profilePowerLaw}
f(\varrho)=Ba^\eta\varrho^\eta
\end{equation}
where  $\eta \geqslant 1$. Important special cases of such punches are a cone with half angle $\theta$ ($\eta = 1$, $B=\frac{1}{\tan(\theta)}$) (fig. \ref{fig:figFourPunchShapes} a) 
\begin{equation}
\label{eq:profileCone}
f(\varrho)=\frac{1}{\tan(\theta)}a\varrho
\end{equation}
and a paraboloid with the radius of curvature $R$ ($\eta=2$ i $B=\frac{1}{2R}$) (fig. \ref{fig:figFourPunchShapes} b)
\begin{equation}
\label{eq:profilePataboloid}
f(\varrho)=\frac{1}{2R}a^2\varrho^2
\end{equation}
A cylindrical punch of the radius $a < 1$ (fig. \ref{fig:figFourPunchShapes} c) can be considered as a limiting case of a power-law-shaped punch, with $\eta \to \infty$. 
The exact profile of a spherical punch of radius $R$ is
\begin{equation}
\label{eq:profileSphere}
f(\varrho)=R-\sqrt{R^2-a^2\varrho^2}
\end{equation}
The paraboloidal profile \eqref{eq:profilePataboloid} is the first non-zero term in the Taylor expansion of the spherical profile \eqref{eq:profileSphere} at the punch apex ($\varrho = 0$). For this reason, the load - indentation relations for a paraboloid are often described as equations for a sphere. However, the case of a punch with the exact spherical profile has been analysed separately in the literature (e.g. \cite{England1962}, \cite{Segedin1957}), so here, we will refer to the punches described by \eqref{eq:profilePataboloid} as paraboloidal.

A blunt conical tip (fig. \ref{fig:figFourPunchShapes} d) can be approximated as a punch shaped as a cone of half-angle $\theta$ with a paraboloidal apex, whose radius of curvature is $R$ \cite{Briscoe1994}. We will refer to it as a blunt cone. The profile of such a tip is given by a piecewise function
\begin{equation}
\label{eq:profileBluntConeParaboloid}
f(\varrho)=\begin{cases}
\frac{a^2\varrho^2}{2R} & \varrho \leqslant \frac{b}{a}\\ \frac{a\varrho}{\tan(\theta)}+\frac{b^2}{2R}-\frac{b}{\tan(\theta)} & \varrho > \frac{b}{a}
\end{cases}
\end{equation}
If $b=\frac{R}{\tan(\theta)}$, then the conical part merges with the apex in such a way that the first derivative of the profile $\dv{f}{\varrho}$ is a continuous function of $\varrho$. We will refer to such punches as smooth blunt cones. Another important special case is a truncated cone, with a flat apex, which can be regarded as a limiting case of \eqref{eq:profileBluntConeParaboloid}, with the apex radius going to infinity. The general problem of the JKR-like contact between a half - space and a sectionally smooth punch, in particular the punch specified by \eqref{eq:profileBluntConeParaboloid}, has been thoroughly analysed in \cite{Maugis1983}. The influence of truncation of a conical punch on  values of Young's modulus for thick rubber samples has been investigated experimentally in \cite{Briscoe1994}. The  hertzian model of contact between a blunt cone and a half-space has been used in several works for analysis of AFM force - distance curves, including studies on endothelial and muscle cells \citep{Mathur2001} as well as on gelatin gels \citep{Uricanu2003}.

\section{Formal expressions for load and contact radius}
\begin{figure}[!t]
	\floatbox[{\capbeside\thisfloatsetup{capbesideposition={left,center},capbesidewidth=6cm}}]{figure}[\FBwidth]
	{\caption{The parameter $\varkappa$ (eq. \ref{eq:alternativeFormsEquilibriumRadius4B}) as a function of the normalized contact radius $\tau = \frac{a}{h}$, for a non-bonded layer. The calculations were carried out for a non-adhesive contact with a power-law-shaped punches of $\eta$ equal to 1 (cone, dotted line), 2 (paraboloid, dashed line) and 3 (solid line). }\label{fig:figCollocationTauILambdaRelationship}}
	{\includegraphics[width=6cm]{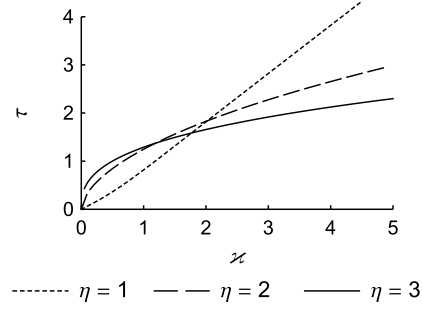}}
\end{figure}

The solution of a Fredholm integral equation of the second kind can be formally represented using the resolvent $H$, which is independent of the free term of the equation. The resolvent corresponding to \eqref{eq:finiteSampleDerZA} depends on the same parameters as the kernel $K$. Thus, $H$ depends on $\tau$ and, in general, on $\nu$, which we will reflect in our notation by writing $H(x,t;\tau,\nu)$. The solution $\chi$ is
\begin{equation}
\label{eq:chiIntermsOfResolvent}
\chi(x;\tau) = \chi^\infty(x)+\frac{1}{\pi}\int_{0}^{1}H(x,t;\tau,\nu)\chi^\infty(t)\dd{t}
\end{equation}
To express load and contact radius in terms of the resolvent $H$, we will introduce the notation
\begin{equation}
\Upsilon^{(\eta)}(\tau;\nu) =\frac{1}{\pi}\int_{0}^{1}H(1,t;\tau,\nu)t^\eta\dd{t}
\end{equation}
\begin{equation}
\label{eq:xiEtaDefinition}
\Xi^{(\eta)}(\tau;\nu) = \frac{1+\eta}{\pi}\int_{0}^{1}\int_{0}^{1}H(x,t;\tau,\nu)t^\eta\dd{t}\dd{x}
\end{equation}
\begin{figure}
	\centering
	\includegraphics{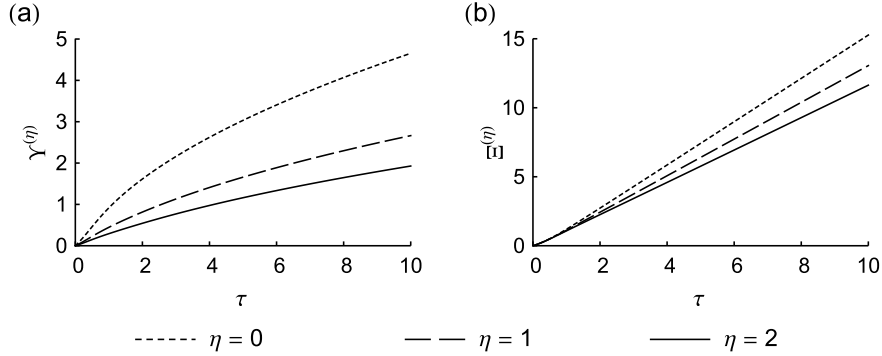}
	\caption{Dimensionless parameters (a) $\Upsilon^{(\eta)}$ and (b) $\Xi^{(\eta)}$ as functions of the ratio $\tau$ of the contact radius $a$ and the layer thickness $h$ for a non-bonded layer, indented with a power-law-shaped punch. The parameters $\Upsilon^{(0)}$ and $\Xi^{(0)}$ (dotted line) appear in equations for any power-law shaped punch, the parameter $\Upsilon^{(1)}$ and  $\Xi^{(1)}$ (dashed line) in equations for a cone, while $\Upsilon^{(2)}$ and $\Xi^{(2)}$ (solid line) in equations for a paraboloid.}
	\label{fig:figCollocationErrorUpsilonEpsilon}
\end{figure}The variation of $\Upsilon^{(\eta)}$ and $\Xi^{(\eta)}$ with $\tau$ for a non-bonded layer is shown in fig. \ref{fig:figCollocationErrorUpsilonEpsilon}. 
\paragraph{Equations for contact radius}
In accordance with \eqref{eq:noAdhesionTauEquilibriumCriterion} and \eqref{eq:chiIntermsOfResolvent}, the equilibrium value of the contact radius $a$ can be calculated from 
\begin{equation}
\label{eq:alternativeFormsEquilibriumRadius2}
0=\chi^\infty(1;a)+\frac{1}{\pi}\int_{0}^{1}H\left(1,t;\tau,\nu \right) \chi^\infty(t;a)\dd{t}
\end{equation}
Combining \eqref{eq:chi_Infinitely_Thick}  and \eqref{eq:alternativeFormsEquilibriumRadius2}, we obtain an equation connecting the equilibrium value of contact radius with indentation depth. For a punch of power-law-shaped profile \eqref{eq:profilePowerLaw}, we arrive at
\begin{equation}
\label{eq:lebedevCollocationInterpolationContactPointC2}
\varkappa= \frac{\sqrt{\pi} \tau^\eta \eulerGamma\mathopen{}\left(\frac{\eta}{2} \right)\mathclose{} }{2\eulerGamma\mathopen{}\left(\frac{1+\eta}{2} \right)\mathclose{} }\left[\frac{1+\Upsilon^{(\eta)}(\tau,\nu)}{1+\Upsilon^{(0)}(\tau,\nu)}  \right]
\end{equation}
where
\begin{equation}
\label{eq:alternativeFormsEquilibriumRadius4B}
\varkappa = \frac{\delta}{\eta B h^\eta}
\end{equation}
The dimension of $B$ is length to the power of $1-\eta$, so that $\varkappa$ is a dimensionless parameter. For a cone, $\varkappa$ is equal to $\frac{\delta\tan(\theta)}{h}$, for a paraboloid $\varkappa$ equals $\frac{R\delta}{h^2}$. The dependence of $\tau$ on $\varkappa$ is shown in fig. \ref{fig:figCollocationTauILambdaRelationship}. 

For a blunt conical punch, $\chi(x)$ can be found by combining \eqref{eq:chi_Infinitely_Thick}, \eqref{eq:profileBluntConeParaboloid} and \eqref{eq:chiIntermsOfResolvent}, which yields
\begin{equation}
\label{eq:chiResolvebrBluntConeParaboloid}
\begin{gathered}
\chi(x) = \frac{2}{\pi}\left(\delta - \frac{a^2 x^2}{R} \right)+\frac{2}{\pi^2}\int_{0}^{1}H\left(x,t;\tau,\nu \right)\left(\delta - \frac{a^2 t^2}{R} \right)\dd{t}+\\+  u\mathopen{}\left(1 - \frac{b}{a} \right)\mathclose{}\frac{2 a^2}{\pi R}\left\lbrace  u\mathopen{}\left(x - \frac{b}{a} \right)\mathclose{}x^2\left[  \sqrt{1-\frac{b^2}{a^2 x^2}} - \frac{R \cot(\theta)}{a x}\acos(\frac{b}{a x}) \right]  \right.  + \\  + \left.\frac{1}{\pi}\int_{\frac{b}{a}}^{1}H\left(x,t;\tau,\nu \right)t^2\left[ \sqrt{1-\frac{b^2}{a^2 t^2}}-\frac{R \cot(\theta)}{a t}\acos(\frac{b}{a t})\right] \dd{t}\right\rbrace 
\end{gathered}
\end{equation}
where $u$ is the unit step function. For a smooth blunt conical punch ($b = R\cot(\theta)$), the equilibrium value of $\tau$ can be found as a root of
\begin{equation}
\begin{gathered}
	0 = \varkappa [1 + \Upsilon^{(0)}(\tau,\nu)] - \tau^2\left[1 + \Upsilon^{(2)}(\tau,\nu) \right] + \\ +  \tau^2 u\mathopen{}\left(1-\frac{\beta}{\tau} \right) \mathclose{}\left[\sqrt{1-\frac{\beta^2}{\tau^2}} - \frac{\beta}{\tau}\acos(\frac{\beta}{\tau}) +  \Theta(\tau,\nu,\beta)\right] 
\end{gathered}
\end{equation}
where 
\begin{equation}
	\Theta(\tau,\nu,\beta) = \frac{1}{\pi}\int_{\frac{\beta}{\tau}}^{1}H(1,t;\tau,\nu)t^2\left[\sqrt{1-\frac{\beta^2}{\tau^2 t^2}} - \frac{\beta}{\tau t}\acos(\frac{\beta}{\tau t}) \right] \dd{t}
\end{equation}

\paragraph{Equations for load}The solution $\chi(x)$ of the Lebedev-Ufliand equation for a thin sample can be expressed by combining the formula \eqref{eq:chiIntermsOfResolvent} for $\chi$ in terms of the resolvent and the solution \eqref{eq:chi_Infinitely_Thick} for the infinitely thick sample. For a power-law-shaped punch
\begin{equation}
\label{eq:chiResolventPowerShaped}
\chi(x)=\frac{2}{\pi}\left[\delta -\frac{\sqrt{\pi}\eta B a^\eta \eulerGamma\mathopen{}\left(\frac{\eta}{2} \right)\mathclose{} }{2\eulerGamma\mathopen{}\left(\frac{1+\eta}{2} \right)\mathclose{} }x^\eta\right] +\frac{2}{\pi^2}\int_{0}^{1}H\left(x,t;\tau,\nu\right)\left[\delta-\frac{\sqrt{\pi}\eta B a^\eta \eulerGamma\mathopen{}\left(\frac{\eta}{2} \right)\mathclose{} }{2\eulerGamma\mathopen{}\left(\frac{1+\eta}{2} \right)\mathclose{} }t^\eta \right] \dd{t}
\end{equation}
To find the equation for load $P$, we substitute \eqref{eq:chiResolventPowerShaped} to the general equation \eqref{eq:axSymInfinitSolJ8} describing the relation between $P$ and $\chi$. After simplification
\begin{equation}
\label{eq:lebedevLoadPowerShaped}
P = \frac{ a E}{1-\nu^2}\left[2\delta\left(1+\Xi^{(0)}(\tau,\nu) \right)- \left(1+\Xi^{(\eta)}(\tau,\nu) \right) \frac{\sqrt{\pi} B a^\eta \eulerGamma\mathopen{}\left(1+\frac{\eta}{2} \right)\mathclose{} }{\eulerGamma\mathopen{}\left(\frac{3+\eta}{2} \right)\mathclose{}}\right] 
\end{equation}
This equation can be rewritten in a dimensionless form, which is more convenient for approximation
\begin{equation}
\label{eq:lebedevReducedLoadPowerShaped}
\Lambda = \tau\left[2\left(1+\Xi^{(0)} \right) \varkappa - \left(1+\Xi^{(\eta)} \right) \frac{\sqrt{\pi} \tau^\eta \eulerGamma\mathopen{}\left(1+\frac{\eta}{2} \right)\mathclose{} }{\eta \eulerGamma\mathopen{}\left(\frac{3+\eta}{2} \right)\mathclose{}}\right] 
\end{equation}
where dimensionless load  $\Lambda$ is defined as $\frac{P(1-\nu^2)}{\eta B E h^{\eta + 1}}$.
If $\tau = 0$, then $\Xi^{(\eta)} = 0$ for any $\eta$. In such a case, \eqref{eq:lebedevLoadPowerShaped} becomes identical with the equation for load necessary to indent a sample thick enough for the effects of the substrate to be negligible 
\begin{equation}
\label{appendix:loadInfiniteThicknessGeneralPowerFunctionTip}
\Lambda = \tau\left[2 \varkappa -  \frac{\sqrt{\pi} \tau^\eta \eulerGamma\mathopen{}\left(1+\frac{\eta}{2} \right)\mathclose{} }{\eta \eulerGamma\mathopen{}\left(\frac{3+\eta}{2} \right)\mathclose{}}\right] 
\end{equation}
If both the effects of the substrate and of adhesion forces can be neglected, the equilibrium value of the contact radius is
\begin{equation}
\label{eq:contactRadiusInfiniteThicknessWithoutAdhesionPowerShape}
\tau=\sqrt[\eta]{\frac{2\varkappa\eulerGamma\mathopen{}\left(\frac{1+\eta}{2} \right)\mathclose{}}{\sqrt{\pi}\eulerGamma\mathopen{}\left(\frac{\eta}{2} \right)\mathclose{}}}
\end{equation}
Substituting \eqref{eq:contactRadiusInfiniteThicknessWithoutAdhesionPowerShape} to \eqref{appendix:loadInfiniteThicknessGeneralPowerFunctionTip}, we obtain
\begin{equation}
\label{eq:loadInfiniteThicknessWithoutAdhesionPowerShape}
\Lambda=\frac{2\eta \varkappa}{(1-\eta)}\sqrt[\eta]{\frac{2\varkappa \eulerGamma\mathopen{}\left(\frac{1+\eta}{2} \right)\mathclose{}}{\sqrt{\pi}\eulerGamma\mathopen{}\left(\frac{\eta}{2} \right)\mathclose{}}}
\end{equation}
This equation was derived by Shtaerman \cite{Shtaerman1939} (for even $\eta$) and Galin \cite{Galin1946} (for any real $\eta \geqslant 1$).

A cone with half-angle $\theta$ is a power-law-shaped punch with $\eta =1$ and $B = \frac{1}{\tan(\theta)}$. Load can be expressed in a dimensionless form as
\begin{equation}
\label{eq:lebedevReduceLoadCone}
\Lambda = \tau\left[ 2\varkappa (1+\Xi^{(0)})-\frac{\pi}{2}\tau(1+\Xi^{(1)})\right]
\end{equation}
where $\Lambda$ is equal to $\frac{P\tan(\theta) (1-\nu^2)}{E h^2}$.
When adhesion forces are absent and the effects of finite sample thickness can be disregarded, the contact radius can be calculated from $\tau = \frac{2}{\pi}\varkappa$. In such a case, \eqref{eq:lebedevReduceLoadCone} assumes the form \citep{Sneddon1965,Love1939}
\begin{equation}
\label{eq:coneSneddon}
\Lambda = \frac{2}{\pi}\varkappa^2
\end{equation}
This equation is often referred to as Sneddon's model.

A paraboloid of radius $R$ is a power-law-shaped punch with $\eta = 2$ i $B = \frac{1}{2R}$. Thus, load can be expressed in a dimensionless form 
\begin{equation}
\label{eq:lebedevReducedLoadParaboloid}
\Lambda = \tau\left[2\varkappa(1+\Xi^{(0)})-\frac{2}{3}\tau^2(1+\Xi^{(2)})\right] 
\end{equation} where $\Lambda = \frac{P R (1-\nu^2)}{E h^3}$. If the contact is non-adhesive and the effects of sample thickness are negligible then $\tau = \sqrt{\varkappa}$, while load can be calculated from Hertz's equation \citep{Hertz1882}
\begin{equation}
\label{eq:paraboloidHertz}
\Lambda = \frac{4}{3}\varkappa^{\frac{3}{2}}
\end{equation} 
For a cylindrical punch, load can be found as
\begin{equation}
\label{eq:lebedevLoadCylinder}
P=\frac{ a E \delta}{1-\nu^2}\left[2\left(1+\Xi^{(0)} \right)\right] 
\end{equation}
Cylindrical indentation requires a different definition of dimensionless load. We will define it as
\begin{equation}
\label{eq:lebedevReducedLoadCylinder}
\Lambda = \frac{P(1-\nu^2)}{2aE\delta}
\end{equation}
The relationship between load and the depth of indentation by a cylindrical punch is linear, regardless of sample thickness. Influence of the rigid substrate on load does not depend on indentation depth. 
In equations \eqref{eq:lebedevLoadPowerShaped} --  \eqref{eq:lebedevLoadCylinder} for load required to indent a thin sample with a power-shaped or cylindrical tip, the effect of the rigid substrate is captured by the parameters $\Xi^{(\eta)}$. For fixed $\eta$, they are functions of $\tau$ (non-bonded) or $\tau$ and $\nu$ (bonded layer). They are not directly dependent on the punch size. Parameters $\Xi^{(0)}$ and $\Xi^{(1)}$ in the equation \eqref{eq:lebedevReduceLoadCone} for load in conical indentation are independent of the cone half-angle. Likewise, $\Xi^{(0)}$ and $\Xi^{(2)}$ in the equation \eqref{eq:lebedevReducedLoadParaboloid} are independent of the radius of the paraboloid.

Substituting \eqref{eq:chiResolvebrBluntConeParaboloid} into \eqref{eq:axSymInfinitSolJ8}, we obtain load required to indent a layer with a blunt conical punch. We will define dimensionless load in the same way as for a paraboloid, by $\Lambda = \frac{P R (1-\nu^2)}{E h^3}$. For a smooth blunt cone ($b = R\cot(\theta)$), we obtain
\begin{equation}
\label{eq:reducedLoadSmoothBluntCone}
\begin{gathered}
\Lambda = \tau \left[2\varkappa(1+\Xi^{(0)}) -\frac{2\tau^2}{3}(1+\Xi^{(2)})+\right. \\ + \left. u\mathopen{}\left(1 - \frac{\beta}{\tau} \right)\mathclose{} \left\lbrace \frac{2\tau^2}{3}\left(1-\frac{\beta^2}{\tau^2} \right)^{\frac{3}{2}}+ \beta\tau\left(\frac{\beta}{\tau}\sqrt{1-\frac{\beta^2}{\tau^2}} - \acos(\frac{\beta}{\tau})\right) + \Psi \right\rbrace  \right] 
\end{gathered}
\end{equation}
where $\beta = \frac{b}{h}$, $\varkappa=\frac{R\delta}{h^2}$,  while $\Psi$ is defined as 
\begin{align}
\label{eq:bluntConePsi}
\Psi(\tau;\nu,\beta) &= \frac{2\tau^2}{\pi}\int_{0}^{1}\int_{\frac{\beta}{\tau}}^{1}H\left(x,t;\tau,\nu \right) t\left[\sqrt{1-\frac{\beta^2}{\tau^2 t^2}}-\frac{\beta}{\tau}\acos(\frac{\beta}{\tau t}) \right] \dd{t}\dd{x}
\end{align}
When  $\frac{\beta}{\tau} > 1$, \eqref{eq:reducedLoadSmoothBluntCone} becomes identical to the equation \eqref{eq:lebedevReducedLoadParaboloid} for a paraboloidal punch. If the effects of sample thickness can be disregarded, then
\begin{equation}
\label{eq:reducedLoadSmoothBluntConeInfiniteThickness}
\begin{gathered}
\Lambda = \tau \left[2\varkappa -\frac{2\tau^2}{3}+\right. \\ + \left. u\mathopen{}\left(1 - \frac{\beta}{\tau} \right)\mathclose{} \left\lbrace \frac{2\tau^2}{3}\left(1-\frac{\beta^2}{\tau^2} \right)^{\frac{3}{2}}+ \beta\tau\left(\frac{\beta}{\tau}\sqrt{1-\frac{\beta^2}{\tau^2}} - \acos(\frac{\beta}{\tau})\right) \right\rbrace  \right] 
\end{gathered}
\end{equation}
Equation \eqref{eq:reducedLoadSmoothBluntConeInfiniteThickness} is a special case of a formula for load required to indent a half-space with a blunt cone in the presence of adhesion forces, derived in \cite{Maugis1983}. The analysis of the effect of the rigid substrate on load required for indentation with a blunt cone is more complex than for power-shaped tips. For a blunt cone, the expression \eqref{eq:reducedLoadSmoothBluntCone} for load features the parameter $\Psi$, which depends on the punch shape through $\beta$.

\section{Approximations of load for conical, paraboloidal and cylindrical tips in the absence of adhesion forces}

If adhesion forces between a power-shaped punch and a layer are absent, the equilibrium value of $\tau$ is a function of $\varkappa$ (non-bonded) or $\varkappa$ and $\nu$ (bonded layer). Thus, dimensionless load for such a punch can also be treated as a function of $\varkappa$ and $\nu$.
We approximated $\Lambda$ with low-degree polynomials $\widehat{\Lambda}$. The goal of approximation was to keep the maximal relative error below $10^{-3}$. Such error is negligible compared to other sources of uncertainty, including the experimental measurement errors and simplifications inherent in the mathematical formulation of the contact problem through linear boundary conditions. In the first step, $\Lambda(\varkappa)$ was expanded as a combination of Chebyshev polynomials of high order (20) using interpolation through Gauss - Chebyshev - Lobatto nodes $\varkappa_i$. These approximations were used as proxies for calculation of less accurate, but simpler expressions. For a non-bonded and a bonded incompressible layer, the final approximants were calculated using the Remez algorithm, available in Mathematica (Wolfram Research, Illinois). For a bonded compressible layer, weighted least squares were used instead. For small $\varkappa$, polynomials in $\varkappa^{\frac{1}{\eta}}$ were used for approximation, as suggested by the asymptotic analysis \citep{Argatov2011}. For medium and large $\varkappa$, the approximants are polynomials in $\varkappa$.

To calculate reduced load in a particular node $\varkappa_i$, we must first find the corresponding equilibrium value of contact radius. Based on the criterion \eqref{eq:noAdhesionTauEquilibriumCriterion}, we searched with Newton's algorithm for a root of $\chi(1;\tau)$, treated as a function of $\tau$. Each evaluation of $\chi(1;\tau)$ requires solving Lebedev - Ufliand equation \eqref{eq:finiteSampleDerZA}, using the Nystr\"{o}m method \cite{Nystrom1930} with a Gauss - Legendre - Radau quadrature that contains a node at 1. The number of nodes necessary to calculate contact radius increases with $\varkappa$ (see Supporting Fig. B1). We used up to $1200$ nodes, depending on $\varkappa$. Once the contact radius corresponding to $\varkappa_i$ is known, reduced load can be calculated from \eqref{eq:lebedevReducedLoadPowerShaped} or \eqref{eq:lebedevReducedLoadCylinder}. The parameters $\Xi^{(\eta)}$ were obtained from 
\begin{equation}
\label{eq:xiFromChi}
\Xi^{(\eta)} = (1+\eta) \int_{0}^{1}[\chi(x) - x^\eta]\dd{x}
\end{equation}
where $\chi$ is the solution of 
\begin{equation}
\label{eq:chiForXi}
\chi(x) = x^\eta + \frac{1}{\pi}\int_{0}^{1}K(x,t)\chi(t)\dd{t} 
\end{equation}
The formula \eqref{eq:xiFromChi} can be derived by combining \eqref{eq:chiIntermsOfResolvent} with $\chi^\infty(x) = x^\eta$ and \eqref{eq:xiEtaDefinition}

The parameter $\Psi$, appearing in the equation \eqref{eq:reducedLoadSmoothBluntCone} for dimensionless load for a smooth, blunt conical punch, was calculated from
\begin{equation}
\label{eq:phiAndPsiFromChi}
\Psi = 2\tau^2 \int_{0}^{1}\left[\chi(x) -\left(x^2\sqrt{1-\frac{\beta^2}{\tau^2 x^2}}-\frac{\beta}{\tau}x\acos(\frac{\beta}{\tau x}) \right) u\mathopen{}\left(x - \frac{\beta}{\tau} \right) \mathclose{}  \right] \dd{x}
\end{equation}
where $u$ is the unit step function, while $\chi$ is  the solution of
\begin{equation}
\label{eq:chiFroPhi}
\chi(x) = \left(x^2\sqrt{1-\frac{\beta^2}{\tau^2 x^2}}-\frac{\beta}{\tau}x\acos(\frac{\beta}{\tau x}) \right) u\mathopen{}\left(x - \frac{\beta}{\tau} \right) \mathclose{} + \frac{1}{\pi}\int_{0}^{1}K(x,t)\chi(t)\dd{t} 
\end{equation}
The formula \eqref{eq:phiAndPsiFromChi} can be obtained in an analogous manner to \eqref{eq:chiForXi}. To solve \eqref{eq:chiForXi}, we used a Gauss - Legendre quadrature $Q_N$ with $N$ nodes. For each node $x_i$, we discretized the equation \eqref{eq:chiForXi} by $Q_N$ and then computed the values of $\chi(x_j)$ ($j=1,\dots,N$) using the Nystr\"{o}m method. The integral of $\chi(x) - x^\eta$ with respect to $x$ was subsequently calculated using the quadrature $Q_N$ and the values $\chi(x_j)$.  In \eqref{eq:chiFroPhi}, the free function  is discontinuous. Thus, we solved it using the Nystr\"{o}m method with two Gauss - Legendre quadratures, $Q^A$ for the interval $[0,\frac{\beta}{\tau}]$ and $Q^B$ in $[\frac{\beta}{\tau}, 1]$.

Solving the Lebedev - Ufliand equation requires multiple evaluations of the function $\Omega$, given by \eqref{eq:finiteSampleDerOmega}, which is a non-elementary cosine transform. To speed up calculations, we expanded $\Omega$ into a series of the Christov functions, as discussed in the \ref{sect:approximationsLebedevUfliandKernel}.

\begin{figure}[!t]
	\centering
	\includegraphics{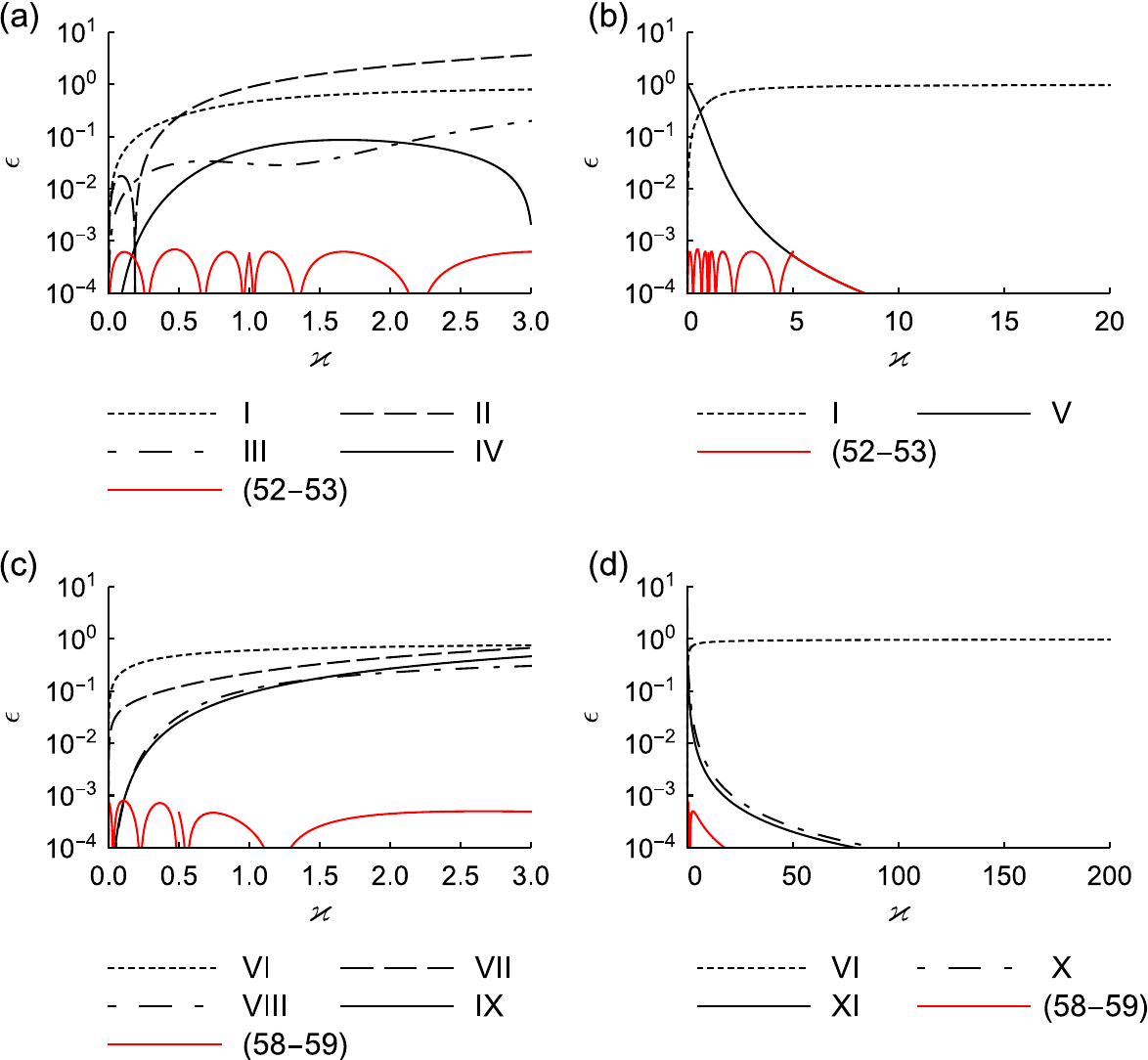}
	\caption{Relative approximation error $\epsilon$ of formulae for force required for indentation of a non-bonded layer with a conical (a, b) or paraboloidal (c, d) punch. Calculations were performed for expressions derived under the assumption of small (a, c) or large (b, d) $\tau$. Red solid lines show the error of the piecewise approximations presented in this work, eq. \eqref{eq:reducedLoadConeA} - \eqref{eq:reducedLoadConeB} in a, b and eq. \eqref{eq:reducedLoadParaboloidA} - \eqref{eq:reducedLoadParaboloidB} in c, d. Dotted lines show the error made when the equations for infinitely thick samples are used (Sneddon's model \citep{Love1939}, \citep{Sneddon1965}, marked as I in a - b, Hertz's model \citep{Hertz1882} marked as VI in c - d). The approximation II was published in \citep{Gavara2012}, III in \citep{Managuli2018}, IV in \citep{Argatov2011}, V in \citep{Yang2003}, VII in \citep{Dimitriadis2002}, VIII in \citep{Vorovich1974}, IX in \citep{Argatov2011}, X in \citep{Aleksandrov1969}, XI in \citep{Yang2003} and \citep{Jaffar1989}.}
	\label{fig:plotErrorsOfLiteratureFormulasConeAndParaboloidLoose}
\end{figure}

\begin{figure}[!t]
	\centering
	\includegraphics{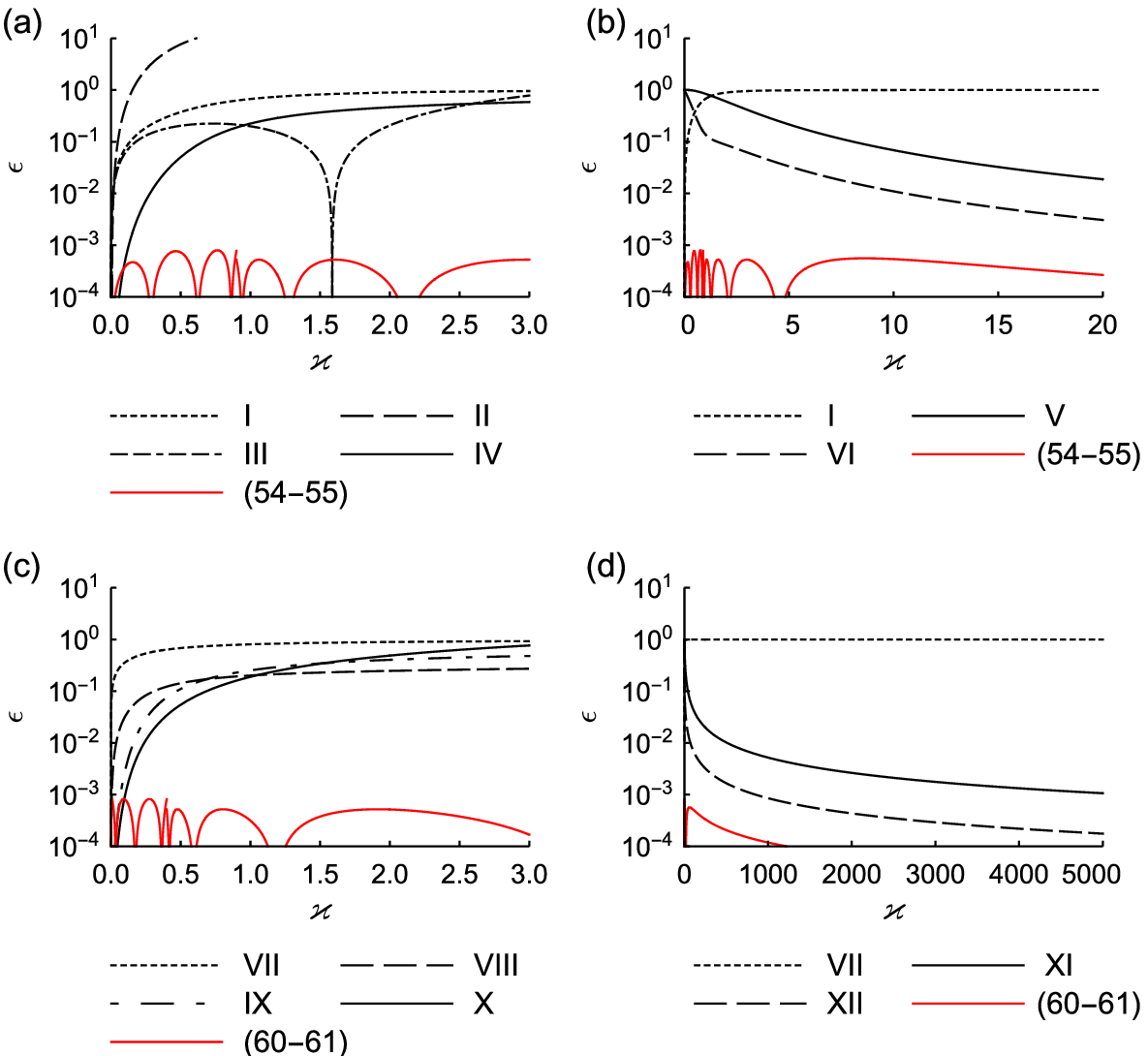}
	\caption{Relative approximation error $\epsilon$ of formulae for force required for indentation of a bonded layer with a conical (a, b) or paraboloidal (c, d) punch. Calculations were performed for expressions derived under the assumption of small (a, c) or large (b, d) $\tau$. Red solid lines show the error of the piecewise approximations presented in this work, eq. \eqref{eq:reducedLoadConeBondedA} - \eqref{eq:reducedLoadConeBondedB} in a, b and \eqref{eq:reducedLoadParaboloidBondedA} \eqref{eq:reducedLoadParaboloidBondedB} in c, d. Dotted lines show the error made when the equations for infinitely thick samples are used (Sneddon's model \citep{Love1939}, \citep{Sneddon1965} marked as I in a, b, Hertz's model \citep{Hertz1882} as VII in c, d). The approximation II was published in \citep{Gavara2012}, III in \citep{Garcia2018}, IV in \citep{Argatov2011} and \citep{Managuli2018}, V in \citep{Yang2003}, VI in \citep{Matthewson1981}, VIII in \citep{Dimitriadis2002}, IX in \citep{Vorovich1974}, X in \citep{Argatov2011}, XI in \citep{Yang2003} and \citep{Jaffar1989}, XII in \citep{Matthewson1981}.}
	\label{fig:plotErrorsOfLiteratureFormulasConeAndParaboloidBonded}
\end{figure}

\paragraph{Conical tips}Dimensionless load required to indent a very thin layer ($\tau \gg 1$) with a conical punch can be calculated using asymptotic expressions, whose leading terms are \cite{Yang2003}
\begin{align}
\label{eq:reducedLoadLooseConeAssymptoticTauLarge}
\Lambda &= \frac{\pi}{3} \varkappa^3\\\label{eq:reducedLoadBondedConeAssymptoticTauLarge}
\Lambda &= \frac{\pi}{80}\left(\frac{3}{2} \right)^5 \varkappa^5
\end{align}
for a non-bonded and an incompressible bonded layer, respectively. For a cone, $\varkappa$ is equal to $\frac{\delta\tan(\theta)}{h}$, in accordance with \eqref{eq:alternativeFormsEquilibriumRadius4B}. Numerical calculations indicate that the relative error of the equation \eqref{eq:reducedLoadLooseConeAssymptoticTauLarge} for a non-bonded layer decreases fast with $\varkappa$, dropping below $10^{-3}$ for $\varkappa > 4.1$ (fig. \ref{fig:plotErrorsOfLiteratureFormulasConeAndParaboloidLoose} b). Thus, the asymptotic formula \eqref{eq:reducedLoadLooseConeAssymptoticTauLarge} was used when $\varkappa > 5$. For smaller $\varkappa$, the reduced load was approximated as
\begin{align}
\label{eq:reducedLoadConeA}
&\widehat{\Lambda}(\varkappa) =\frac{2}{\pi}\varkappa^2 \left(1 + 0.461\varkappa + 
0.346\varkappa^2 + 0.0484\varkappa^3\right)  \enspace \varkappa \leqslant 1 \\&\widehat{\Lambda}(\varkappa) = 0.0859 + 0.103 \varkappa - 0.0647\varkappa^2 + 1.057\varkappa^3\quad  1 < \varkappa \leqslant 5
\label{eq:reducedLoadConeB}
\end{align} The relative error of the asymptotic expression \eqref{eq:reducedLoadBondedConeAssymptoticTauLarge} for a bonded layer decreases much slower than the error of the corresponding expression \eqref{eq:reducedLoadLooseConeAssymptoticTauLarge} for a non-bonded layer (fig. \ref{fig:plotErrorsOfLiteratureFormulasConeAndParaboloidBonded} b). It drops below $10^{-3}$ only when $\varkappa > 88.9$. Thus, we calculated approximations in a much wider interval than in the case of non-bonded layer
\begin{align}
\label{eq:reducedLoadConeBondedA}
&\widehat{\Lambda}(\varkappa) =\frac{2}{\pi}\varkappa^2 \left(1 +0.715\varkappa + 
0.609\varkappa^2 + 0.735\varkappa^3\right) \qquad \qquad \qquad \enspace \varkappa \leqslant 0.9 \\&\widehat{\Lambda}(\varkappa) = -0.265  +  1.225\varkappa - 1.651\varkappa^2 + 2.332\varkappa^3+\frac{\pi}{80}\left(\frac{3}{2} \right)^5 \varkappa^5\quad  0.9 < \varkappa \label{eq:reducedLoadConeBondedB}
\end{align}
For $\varkappa > 89$, the equation \eqref{eq:reducedLoadBondedConeAssymptoticTauLarge} can be used. The relative errors of the derived approximations for load required to indent a layer with a conical tip are shown in fig. \ref{fig:plotErrorsOfLiteratureFormulasConeAndParaboloidLoose} a, b (non - bonded) and \ref{fig:plotErrorsOfLiteratureFormulasConeAndParaboloidBonded} a, b (bonded layer).
\paragraph{Paraboloidal tips}When $\tau$ is large, dimensionless load for paraboloidal indentation can be approximated as \cite{Maltis2005}, \cite{Jaffar1989, Yang2003, Malits2006}
\begin{align}
\label{eq:reducedLoadLooseParaboloidAssymptoticTauLarge}
	\Lambda &= \pi \varkappa^2\\ \label{eq:reducedLoadBondedParaboloidAssymptoticTauLarge}
	\Lambda &= \frac{\pi}{2} \varkappa^3
\end{align}
for a non-bonded and an incompressible bonded layer, respectively. For a paraboloid, $\varkappa$ is equal to $\frac{R \delta}{h^2}$, in accordance with \eqref{eq:alternativeFormsEquilibriumRadius4B}. Our calculations for a non-bonded layer indicate that the relative error of \eqref{eq:reducedLoadLooseParaboloidAssymptoticTauLarge} is less than $10^{-3}$ when $\varkappa > 17.1$. The interval of approximation was split into three regions. In the first two of them
\begin{align}
\label{eq:reducedLoadParaboloidA}
&\widehat{\Lambda}(\varkappa) =\frac{4}{3}\varkappa^{\frac{3}{2}} \left(1 + 0.722\varkappa^{\frac{1}{2}} + 
0.822\varkappa\right)  \enspace \varkappa \leqslant 0.5 \\&\widehat{\Lambda}(\varkappa) =-0.0633 + 0.260 \varkappa^{\frac{1}{2}}+\pi\varkappa^2 \qquad  0.5 < \varkappa \leqslant 450
\label{eq:reducedLoadParaboloidB}
\end{align}
In the third region of $\varkappa > 450$, the asymptotic formula \eqref{eq:reducedLoadLooseParaboloidAssymptoticTauLarge} was used. In the case of a bonded layer, the relative error of the asymptotic expression for very thin layers (eq. \ref{eq:reducedLoadLooseConeAssymptoticTauLarge}) drops below $10^{-3}$ only when $\varkappa >5313$. The corresponding piecewise approximations of $\Lambda$ are
\begin{align}
\label{eq:reducedLoadParaboloidBondedA}
&\widehat{\Lambda}(\varkappa) =\frac{4}{3}\varkappa^\frac{3}{2} \left(1 +1.105\varkappa^{\frac{1}{2}} + 
1.607\varkappa + 1.602\varkappa^\frac{3}{2}\right) \qquad \qquad \qquad \enspace \varkappa \leqslant 0.4 \\&\widehat{\Lambda}(\varkappa) = 0.616 - 3.114\varkappa^{\frac{1}{2}}+ 6.693\varkappa -7.170 \varkappa^\frac{3}{2}+ 8.228\varkappa^2+\frac{\pi}{2} \varkappa^3\;\;  0.4 < \varkappa \label{eq:reducedLoadParaboloidBondedB}
\end{align}
We calculated the relative error of approximation \eqref{eq:reducedLoadParaboloidBondedA} - \eqref{eq:reducedLoadParaboloidBondedB} for
$\varkappa \leqslant 20000$. In this range, the error is negligible. The errors of the proposed approximations for paraboloidal tips are shown in fig. \ref{fig:plotErrorsOfLiteratureFormulasConeAndParaboloidLoose} c, d (non-bonded) and fig. \ref{fig:plotErrorsOfLiteratureFormulasConeAndParaboloidBonded} c, d (bonded layer).

\begin{figure}[!t]
	\centering
	\includegraphics{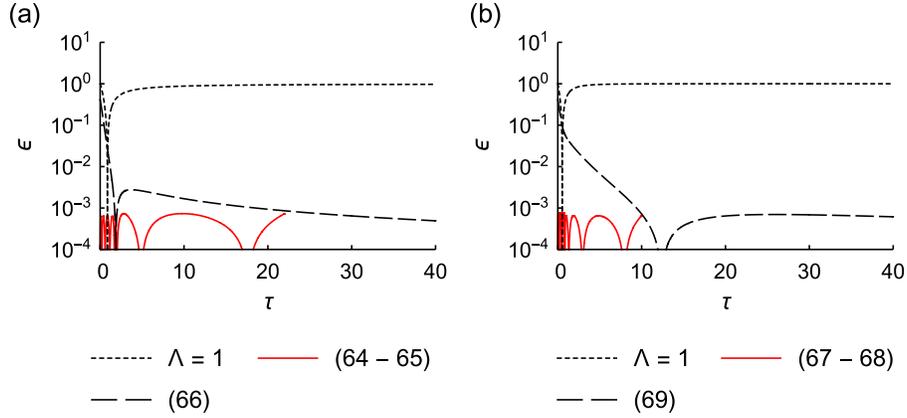}
	\caption{Relative error $\epsilon$ of formulae for dimensionless load $\Lambda$ required a non-bonded (a) or bonded (b) layer with a cylindrical punch. Red solid lines show the error of the piecewise approximations, \eqref{eq:cylindricalTipLoadLooseApproximationsStart} - \eqref{eq:cylindricalTipLoadLooseApproximationsEnd} in a and \eqref{eq:cylindricalTipLoadBondedApproximationsStart} - \eqref{eq:cylindricalTipLoadBondedApproximationsEnd} in b. Dashed lines show the error of formulae designed for large $\varkappa$, which are \eqref{eq:xiApproximationsMalitsLoose} in a and \eqref{eq:malitsIncompressibleBonded} in b. Dotted lines show the error made when the equation $\Lambda(\tau) = 1$ is used, which neglects the effect of the rigid substrate.}
	\label{fig:plotReducedForceConeAndParaboloidChebyshevLowDegreeApproximations}
\end{figure}

\paragraph{Cylindrical tips}
The leading terms of the asymptotic expansion of the dimensionless load $\Lambda$ required for cylindrical indentation of a very thin layer ($\tau \gg 1$) are \citep{Barber1990}
\begin{align}
\Lambda &= \frac{\pi}{2}\tau\\
\Lambda &= \frac{3\pi}{64}\tau^3
\end{align}
for a non-bonded and an incompressible bonded layer, respectively. We calculated approximations of  $\Lambda$ separately in two intervals. For a non-bonded layer, we obtained
\begin{align}
\label{eq:cylindricalTipLoadLooseApproximationsStart}
&\widehat{\Lambda}(\tau) = 1+0.725\tau + 0.697\tau^2 -0.259\tau^3 + 0.0347 \tau^4  \enspace \tau \leqslant 1.8 \\&\widehat{\Lambda}(\tau) =0.602 +1.562\tau + 
0.000371\tau^2 \quad   1.8 < \tau \leqslant 22\label{eq:cylindricalTipLoadLooseApproximationsEnd}
\end{align}
For $\tau > 22$, we used the asymptotic expansion derived in \citep{Maltis2005}, truncated to the first two terms
\begin{align}
\label{eq:xiApproximationsMalitsLoose}
\Lambda &= \frac{\pi}{2}\tau + 0.589
\end{align}
For a bonded layer, the piecewise approximations take the form of
\begin{align}
\label{eq:cylindricalTipLoadBondedApproximationsStart}
&\widehat{\Lambda}(\tau) = 1+ 1.101\tau + 1.518\tau^2 + 0.0207\tau^3  \enspace \tau \leqslant 0.8 \\&\widehat{\Lambda}(\tau) = 0.713 +1.893\tau + 
0.869\tau^2 + 0.146 \tau^3 \quad   0.8 < \tau \leqslant 10\label{eq:cylindricalTipLoadBondedApproximationsEnd}
\end{align}
If $\tau > 10$, then $\Lambda$ can be calculated with negligible error, using the asymptotic expansion proposed in \citep{Malits2006}, truncated to
\begin{equation}
\label{eq:malitsIncompressibleBonded}
	\widehat{\Lambda} = \frac{1}{4\beta^3}\left(1+1.924\beta+1.924\beta^2-2.520\beta^3 \right) 
\end{equation}
where $\beta = \frac{1}{1+\left(\frac{3\pi}{16} \right)^{\frac{1}{3}}\tau}$. The relative error of piecewise approximations of reduced load $\Lambda$ for a non-bonded and a bonded layer is shown in fig. \ref{fig:plotReducedForceConeAndParaboloidChebyshevLowDegreeApproximations} a, b.

\begin{figure}[!t]
	\centering
	\includegraphics{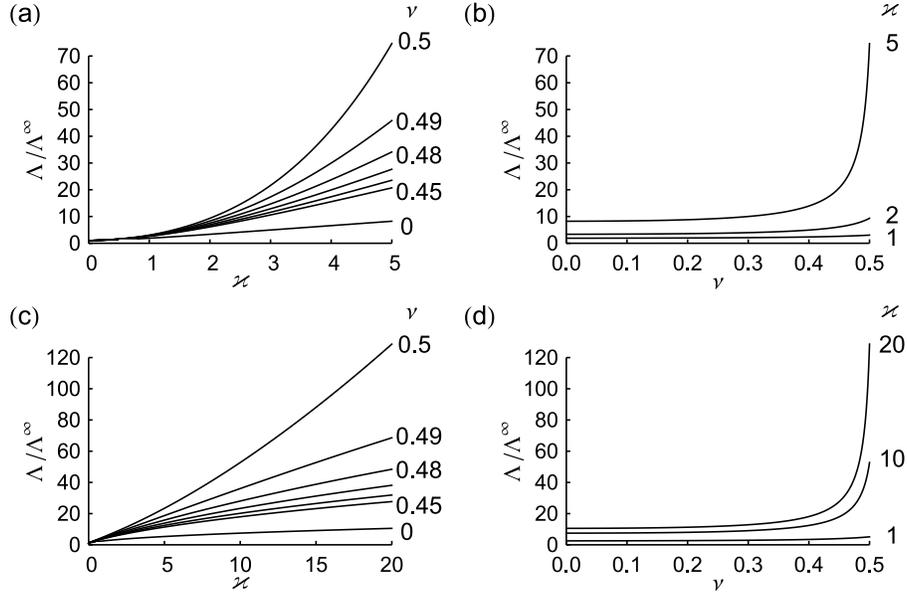}
	\caption{The dependence of the dimensionless load $\Lambda$ required to indent a bonded layer on its Poisson's ratio, calculated for conical (a, b) and paraboloidal (c, d) punches. The values of $\Lambda$ were normalized by $\Lambda^\infty$, which is dimensionless load calculated under the assumption that the sample thickness does not influence the parameters of contact. For a conical punch, $\Lambda^\infty = \frac{2}{\pi}\varkappa^2$ (Sneddon's model, eq. \eqref{eq:coneSneddon}), for paraboloidal $\Lambda^\infty = \frac{4}{3}\varkappa^{\frac{3}{2}}$ (Hertz's model, eq. \eqref{eq:paraboloidHertz}). Note that the vertical axis range in (a, b) differs from that in (c,d).}
	\label{fig:plotReducedLoaBondedImportanceOfPoissonRatio}
\end{figure}

The equations for conical, paraboloidal and cylindrical tips presented above do not describe indentation of a bonded, compressible layer. In such a case, dimensionless load $\Lambda$ is also a function of Poisson's ratio of the sample $\nu$. Numerical results indicate that the influence of $\nu$ on $\Lambda$ grows with $\varkappa$. In particular, when the layer is very thin and nearly incompressible (i.e. with $\nu$ close to 0.5), small increments of $\nu$ leads to a large increase of $\Lambda$ (fig. \ref{fig:plotReducedLoaBondedImportanceOfPoissonRatio}). To approximate the relationship between $\Lambda$ and $\varkappa$ with the relative error below $10^{-3}$, we had to split the domain $\varkappa$ - $\nu$ ($0 \leqslant \nu \leqslant 0.5$, $0 \leqslant \varkappa $) into multiple regions. As they are numerous, these equations are not included in this paper, but they can be found in the publicly available source code of AtomicJ. 

\begin{figure}[!h]
	\floatbox[{\capbeside\thisfloatsetup{capbesideposition={left,center},capbesidewidth=6cm}}]{figure}[\FBwidth]
	{\caption{The relative error $\epsilon$ of approximate equations for the equilibrium radius of contact between a smooth blunt conical tip and an infinitely thick sample. The dotted line shows the error made when the  presence of the round apex is neglected, i.e. the equation $a = \frac{2}{\pi}\delta \tan(\theta)$ for a conical punch is used. The relative error of the Lagrange inversion expansion \eqref{eq:smoothBluntConeInfinitelyThickContactRadiusApproximationLagrangeSmall} is smaller than $1\%$ if $\frac{R\delta}{b^2} \leqslant 2$ (dashed line), while the error of \eqref{eq:smoothBluntConeInfinitelyThickContactRadiusApproximationLagrangeLarge} is below $1\%$ for $\frac{R\delta}{b^2} > 2$ (dash-dotted line).The error of the piecewise numerical approximation \eqref{eq:smoothBluntConeInfinitelyThickContactRadiusApproximationNumerical} is below $0.06\%$ for any $\frac{R\delta}{b^2} \geqslant 1$ (red solid line). When $\frac{R\delta}{b^2}$ is smaller than 1, the exact formula $a = \sqrt{R\delta}$ for a paraboloidal punch should be used. }\label{fig:plotSmoothBluntConeErrorInfiniteThickness}}
	{\includegraphics[width=6cm]{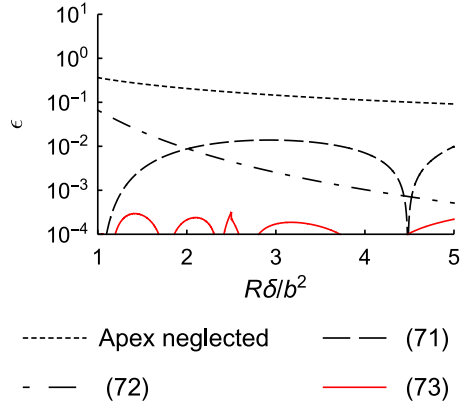}}
\end{figure}

\paragraph{Blunt conical tips}
Dimensionless load required to indent an infinitely thick sample with a smooth blunt cone is an explicit function of $a$, $b$ and $R\delta$. However, the relationship between the equilibrium value of $\frac{a}{b}$ and $\frac{R\delta}{b^2}$ for $\frac{a}{b} < 1$ is given implicitly by
\begin{equation}
\label{eq:smoothBluntConeInfinitelyThickContactRadiusEquation}
\begin{gathered}
\frac{R\delta}{b^2} =  \frac{a^2}{b^2} \left[1-\sqrt{1-\frac{b^2}{a^2}} + \frac{b}{a}\acos(\frac{b}{a})\right] 
\end{gathered}
\end{equation}
If we denote the right hand side of \eqref{eq:smoothBluntConeInfinitelyThickContactRadiusEquation} by $h(\frac{a}{b})$, then solving this equation amounts to finding the inverse function $h^{-1}(\frac{a}{b})$. A series expansion of $h^{-1}$ in a neighbourhood of $h(1) = 1$ can be found using the extended Lagrange inversion theorem (\citep{Olver2010}, p. 21, eq. 1.10.15 - 1.10.16) after substitution $y = \sqrt{\frac{a}{b} - 1}$. Truncating the expansion to the first three non-zero terms, we obtain an expression for $\tau$ valid when $\frac{R\delta}{b^2}$ is sufficiently close to $1$
\begin{equation}
\label{eq:smoothBluntConeInfinitelyThickContactRadiusApproximationLagrangeSmall}
	a = \left[\frac{\sqrt{R\delta - b^2}}{\sqrt{2}\sqrt{b}} +\frac{R\delta - b^2}{6\sqrt{2}b^{\frac{3}{2}}} - \frac{(R\delta - b^2)^{\frac{3}{2}}}{18\sqrt{2}b^{\frac{5}{2}}} + O\mathopen{}\left((R\delta - b^2)^{2} \right)\mathclose{} \right] ^2 +b 
\end{equation}
Note that due to the presence of square in \eqref{eq:smoothBluntConeInfinitelyThickContactRadiusApproximationLagrangeSmall}, the error of $\tau$ is ${O((R\delta-b^2)^{\frac{5}{2}})}$. Using a similar expansion for $\frac{a}{b} \gg 1$, we obtain 
\begin{equation}
\label{eq:smoothBluntConeInfinitelyThickContactRadiusApproximationLagrangeLarge}
a = \left[ \sqrt{\frac{2}{\pi b}}\sqrt{R\delta}  - \frac{(\pi - 1)b^{\frac{3}{2}}}{2\sqrt{2\pi}} \frac{1}{\sqrt{R\delta}} - \frac{(\pi -1)^2 b^{\frac{7}{2}}}{16\sqrt{2\pi}}\frac{1}{(R\delta)^{\frac{3}{2}}} + O\mathopen{}\left(\frac{1}{(R\delta)^{\frac{5}{2}}} \right) \mathclose{}\right] ^2 + b
\end{equation}
provided that $\frac{R\delta}{b^2}$ is sufficiently large. The relative error of approximations \eqref{eq:smoothBluntConeInfinitelyThickContactRadiusApproximationLagrangeSmall} and \eqref{eq:smoothBluntConeInfinitelyThickContactRadiusApproximationLagrangeLarge} is a function of $\frac{R\delta}{b^2}$. Numerical calculations indicate that when \eqref{eq:smoothBluntConeInfinitelyThickContactRadiusApproximationLagrangeSmall} is used for $\frac{R\delta}{b^2} \leqslant 2$ and \eqref{eq:smoothBluntConeInfinitelyThickContactRadiusApproximationLagrangeLarge} for $\frac{R\delta}{b^2} > 2$, the relative error of $a$ never exceeds $1\%$ (fig. \ref{fig:plotSmoothBluntConeErrorInfiniteThickness}). This error is already negligible, but we can reduce it by more than an order of magnitude without introducing more terms. Using numerical methods to find a piecewise approximation with the same general form as that of expansions \eqref{eq:smoothBluntConeInfinitelyThickContactRadiusApproximationLagrangeSmall} and \eqref{eq:smoothBluntConeInfinitelyThickContactRadiusApproximationLagrangeLarge}, we obtain
\begin{equation}
\label{eq:smoothBluntConeInfinitelyThickContactRadiusApproximationNumerical}
a \approx \begin{cases}
\left( 0.712\frac{\sqrt{R\delta - b^2}}{\sqrt{b}} +0.0999\frac{R\delta - b^2}{b^{\frac{3}{2}}} - 0.0350\frac{(R\delta - b^2)^{\frac{3}{2}}}{b^{\frac{5}{2}}}\right) ^2 +b & 1 \leqslant \frac{R\delta}{b^2} \leqslant 2.5\\ \left( \sqrt{\frac{2}{\pi }}\sqrt{\frac{R\delta}{b}}  - 0.422 \frac{b^{\frac{3}{2}}}{\sqrt{R\delta}} - 0.144\frac{b^{\frac{7}{2}}}{(R\delta)^{\frac{3}{2}}} \right) ^2 + b & \frac{R\delta}{b^2} > 2.5
\end{cases}
\end{equation}
When only the round apex of the punch is in contact with the sample, i.e. when $\frac{R\delta}{b^2} < 1$, the equation $a = \sqrt{R\delta}$ for a paraboloidal punch should be used.

The asymptotic expansions of load required to indent a thin layer have apparently not been discussed in the literature. We obtained an expansion valid for small $\tau$ using the general method presented in \citep{England1962} 
\begin{equation}
\label{eq:englandConeParaboloidalApexLoad}
\Lambda=\frac{1}{3}\left[ 4\tau_0^3+\tau_0(\beta^2 - 4\tau_0^2)\sqrt{1-\frac{\beta^2}{\tau_0^2}}+3\tau_0^2 \beta\acos(\frac{\beta}{\tau_0}) \right]  
\end{equation}
where $\Lambda = \frac{P R (1-\nu^2)}{E h^3}$, while $\tau_0$ can be calculated as a solution of
\begin{equation}
\label{eq:englandConeParaboloidalApexIndentation}
\begin{gathered}
\frac{\varkappa}{\beta^2} \approx \frac{\tau_0}{\beta}\left[ \frac{\tau_0}{\beta}+\acos(\frac{\beta}{\tau_0})-\frac{\tau_0}{\beta} \sqrt{1-\frac{\beta^2}{\tau_0^2}}\right]  +\\+\frac{K_0}{6}\tau_0\left[\left(1-4\frac{\tau_0^2}{\beta^2}  \right) \sqrt{1-\frac{\beta^2}{\tau_0^2}}+3\acos(\frac{\beta}{\tau_0}) \frac{\tau_0}{\beta}+4\frac{\tau_0^2}{\beta^2} \right] +\\+\frac{K_1}{180}\tau_0\beta^2\left[\left(  6 +13 \frac{\tau_0^2}{\beta^2}-64\frac{\tau_0^4}{\beta^4}\right)  \sqrt{1-\frac{\beta^2}{\tau_0^2}} + 45\acos(\frac{\beta}{\tau_0}) \frac{\tau_0^3}{\beta^3}+64\frac{\tau_0^4}{\beta^4} \right]  +\\+
\frac{K_2}{630}\tau_0\beta^4\left[\left(  8 +46 \frac{\tau_0^2}{\beta^2}+45 \frac{\tau_0^4}{\beta^4}-288\frac{\tau_0^6 }{\beta^6}\right)  \sqrt{1-\frac{\beta^2}{\tau_0^2}} + 189\acos(\frac{\beta}{\tau_0}) \frac{\tau_0^5}{\beta^5}+288\frac{\tau_0^6}{\beta^6} \right] 
\end{gathered} 
\end{equation}
\begin{table}[!b]
	\caption{Values of coefficients $K_0$ -- $K_3$, defined in \eqref{eq:asymptoticKi}. These coefficients appear in the equations \eqref{eq:englandConeParaboloidalApexIndentation} - \eqref{eq:englandConeParaboloidalApexContactRadius} for a blunt paraboloidal punch indenting a layer.}
	\label{table:kiCoefficientsValuesAsymptotic}
	\begin{tabular}{ c c c c c c c c}
		\hline
		&\specialcell[c]{Loose\\layer}& \multicolumn{6}{c}{Bonded layer}\\ & &$\nu=0$& $\nu=0.1$ &$\nu=0.2$& $\nu=0.3$& $\nu=0.4$ & $\nu=0.5$\\
		\hline
		$K_0$ & -0.7433 & -0.7710 & -0.7904 & -0.8229 & -0.8765 & -0.9668 & -1.1270\\
		$K_1$ & 0.5034 & 0.5980 & 0.6406 & 0.7035 & 0.7990 & 0.9518 & 1.2195\\
		$K_2$ & -0.2184 & -0.3077 & -0.3391 & -0.3837 & -0.4500 & -0.5552 & -0.7425\\
		$K_3$ & 0.0772  & 0.1293 & 0.1449 & 0.1667 & 0.1988 & 0.2498 & 0.3415\\
		\hline
	\end{tabular}
\end{table}To find the load corresponding to a particular dimensionless indentation depth $\varkappa$, we first solve numerically \eqref{eq:englandConeParaboloidalApexIndentation} for $\tau_0$, and then substitute it into \eqref{eq:englandConeParaboloidalApexLoad}. Note that in general the parameter $\tau_0$ is not equal to the equilibrium value $\tau$ of dimensionless contact radius and should be treated as a parameter that has to be calculated numerically. The approximate equilibrium value of the dimensionless contact radius $\tau$ can be found from
\begin{equation}
\label{eq:englandConeParaboloidalApexContactRadius}
\begin{gathered}
	\tau \approx \tau_0 -\frac{K_1\tau_0^4\left[4 + \left(\frac{\beta^2}{\tau_0^2} -4 \right) \sqrt{1-\frac{\beta^2}{\tau_0^2}}+ 3\frac{\beta}{\tau_0}\acos(\frac{\beta}{\tau_0}) \right] }{9\left[2\left(1-\sqrt{1-\frac{\beta^2}{\tau_0^2}} \right) +\frac{\beta}{\tau_0}\acos(\frac{\beta}{\tau_0}) \right] } +\\+ \frac{K_2\tau_0^6}{15}\left[\frac{16+\left(3\frac{\beta^2}{\tau_0^2}+2\frac{\beta^4}{\tau_0^4}-16 \right) \sqrt{1-\frac{\beta^2}{\tau_0^2}}+11\frac{\beta}{\tau_0}\acos(\frac{\beta}{\tau_0})}{2\left(\sqrt{1-\frac{\beta}{\tau_0}} -1\right)-\frac{\beta}{\tau_0}\acos(\frac{\beta}{\tau_0}) } \right] 
\end{gathered}
\end{equation}
\begin{figure}[!t]
	\centering
	\includegraphics{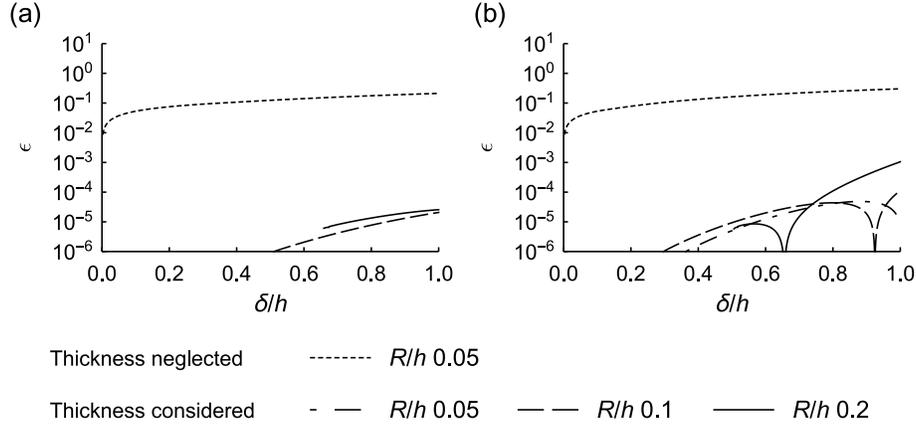}
	\caption{Relative error $\epsilon$ of asymptotic approximations \eqref{eq:englandConeParaboloidalApexLoad} -- \eqref{eq:englandConeParaboloidalApexIndentation} of dimensionless load $\Lambda$ required to indent a thin, non-bonded layer with a smooth blunt conical punch. Calculations were performed for punches with half-angle of (a) $20^\circ$ and (b) $30^\circ$, with radius of the apex curvature equal to 0.05 (dash-dotted line), 0.1 (dashed line) or 0.2 (solid line) of layer thickness. Dotted lines show the error made when the equation \eqref{eq:reducedLoadSmoothBluntConeInfiniteThickness} is used (shown for $R/h = 0.4$, the error grows with $R/h$). This equation neglects the effect of sample thickness. }
	\label{fig:bluntConeRelativeErrorTan20DegAnd30DDeg}
\end{figure}The coefficients $K_i$ are defined as even moments of the weight function $\omega$
\begin{equation}
\label{eq:asymptoticKi}
K_i=-\frac{2(-1)^i}{\pi(2i)!}\int_{0}^{\infty}\omega(t \tau;\tau)t^{2i}\dd{t}
\end{equation}

For a layer bonded to the rigid substrate, $K_i$ is a function of Poisson's ratio $\nu$. Values of $K_i$ for selected $\nu$ are shown in table \ref{table:kiCoefficientsValuesAsymptotic}. Approximate values of $K_0$ -- $K_3$ for a bonded layer ($0 \leqslant v \leqslant 0.5$) can be calculated using the following least-squares approximations
\begin{align}
\label{eq:assymptoticK0Approximation}
K_0&=\left(-0.77094 +2.16403\nu-3.39663\nu^2+3.13918\nu^3 -1.69195\nu^4 \right)\euler^{3\nu}\\
\label{eq:assymptoticK1Approximation}
K_1&=\left(0.59774-2.02688\nu+3.84145\nu^2-4.21612\nu^3+2.35827\nu^4 \right)\euler^{4\nu} 
\\
\label{eq:assymptoticK2Approximation}
K_2&=\left(-0.30770 + 0.96407\nu-1.78726\nu^2+1.99104\nu^3-1.22965\nu^4\right)\euler^{4\nu}\\
\label{eq:assymptoticK3Approximation}
K_3&=\left( 0.12934-0.38527 \nu +0.70317 \nu^2 -0.79353\nu^3 + 0.52607\nu^4 \right)\euler^{4\nu}
\end{align}
The maximal relative error of \eqref{eq:assymptoticK0Approximation} -- \eqref{eq:assymptoticK3Approximation} in the interval $\nu \in [0,0.5]$ does not exceed $0.05\%$. 

Numerical calculations indicate that the approximations \eqref{eq:englandConeParaboloidalApexLoad} -- \eqref{eq:englandConeParaboloidalApexContactRadius} are accurate when normalized contact radius $\tau$ is sufficiently small. The relative error due to approximation is below $1\%$ for any indentation depth provided that $R/h \leqslant 0.3$ and $\theta \leqslant 34.5^\circ$ (fig. \ref{fig:bluntConeRelativeErrorTan20DegAnd30DDeg}). These conditions hold in typical AFM experiments.

When radius of contact between a blunt cone and a layer is smaller than $b$, the sample interacts only with the paraboloidal apex. In such a case, load should be calculated from equations for a paraboloidal punch. An approximation of the relation between $\tau$ and $\varkappa$ for a paraboloidal punch can facilitate the choice of the appropriate model. We obtained piecewise formulae describing $\tau(\varkappa)$ for a paraboloid using the Remez algorithm. For a non-bonded layer, approximate values  of $\tau$, denoted as $\widehat{\tau}$, can be calculated from
\begin{align}
\label{eq:paraboloidTausVsVarKappaA}
&\widehat{\tau}(\varkappa) = \sqrt{\varkappa} \left(1 + 0.244\sqrt{\varkappa} + 
0.212\varkappa - 0.303\varkappa^\frac{3}{2} + 0.0954\varkappa^2\right)  \enspace \varkappa \leqslant 2 \\&\widehat{\tau}(\varkappa) = -0.172 + 1.413 \sqrt{\varkappa} \hskip14em\relax  1 < \varkappa \leqslant 452
\label{eq:paraboloidTausVsVarKappaB}
\end{align}
For a bonded incompressible layer ($\nu = 0.5$), the approximations are
\begin{align}
\label{eq:paraboloidTausVsVarKappaABonded}
&\widehat{\tau}(\varkappa) =\sqrt{\varkappa} \left(1 + 0.359\varkappa^{\frac{1}{2}} + 
0.556\varkappa - 
0.789 \varkappa^{\frac{3}{2}} + 
0.283 \varkappa^2 \right)  \enspace \varkappa \leqslant 1.5 \\&\widehat{\tau}(\varkappa) = -0.224 + 1.552\varkappa^{\frac{1}{2}} + 
0.0789\varkappa - 
0.00520\varkappa^{\frac{3}{2}} \qquad 1.5 < \varkappa \leqslant 40\\&\widehat{\tau}(\varkappa) = -1.106 + 1.983\varkappa^{\frac{1}{2}}  \hspace{4cm}40 < \varkappa \leqslant 245.69
\label{eq:paraboloidTausVsVarKappaBBonded}
\end{align}
where $\varkappa = \frac{R\delta}{h^2}$. For larger values of $\varkappa$, equilibrium values of radius of contact radius can be calculated from equations presented in \cite{Yang2003}, \cite{Jaffar1989}, 
\begin{align}
\label{eq:paraboloidTauLooseVeryThinLayer}
\tau &= \sqrt{2}\varkappa^{\frac{1}{2}}\quad \textrm{(non-bonded)} \\
\tau &= 2 \varkappa^{\frac{1}{2}} \quad \textrm{(bonded, $\nu = 0.5$)} 
\label{eq:paraboloidTauBondedVeryThinLayer}
\end{align}
The relative error of piecewise approximations \eqref{eq:paraboloidTausVsVarKappaA} -- \eqref{eq:paraboloidTausVsVarKappaB} and \eqref{eq:paraboloidTausVsVarKappaABonded} -- \eqref{eq:paraboloidTausVsVarKappaBBonded} does not exceed $10^{-3}$ within the range of approximation (Supporting Figure \ref{fig:plotsTauErrorsParaboloidLooseAndBoundedIncompressible}). 

\section{Discussion}
In this paper, we presented low-degree piecewise polynomial approximations of load required to indent a linear-elastic, isotropic layer with a conical, paraboloidal or cylindrical punch. The equations have been obtained under the assumption that the contact between the punch and the layer is frictionless, while the layer can either freely slip on the rigid substrate (the non-bonded case) or it is bonded to the substrate. Through combination of numerical techniques and known asymptotic formulae for very thin layers, we obtained equations that describe layers of arbitrary thickness. Comparison of the approximate formulae with our numerical solutions of the Lebedev - Ufliand integral equation \citep{Lebedev1958} indicates that their relative error is smaller than $10^{-3}$. Thus, it is negligible compared to other contributions to the error of the values of Young's modulus calculated from force - distance curves, for example uncertainty of cantilever's spring constant, inhomogeneity of the sample, deviations from the assumptions of the linear elasticity or the effect of lateral displacements of the sample material, which are not accounted for by the hertzian conditions \eqref{eq:axiSymContactBoundaryA} - \eqref{eq:axiSymContactBoundaryC}. The equations presented in this paper have been implemented in the new version of AtomicJ, freely available through the SourceForge platform. AtomicJ allows for concurrent processing of multiple force distance curves and force - volume recordings. For analysis of force curves recorded on a thin sample, the thickness can be automatically read from a user-specified topographical image. 

Formulae for load suitable for layers of arbitrary thickness can mitigate problems that arise during implementation of popular procedures for analysis of force - distance curves. To calculate Young's modulus, it is necessary to identify the point of initial contact $(z_0,P_0)$ between the tip and the sample. A force - distance curve consists of the off - contact ($z < z_0$) and in-contact ($z \geqslant z_0$) regions. Using standard procedures of the contact point identification, it is necessary to assume a theoretical model of the load - indentation relation suitable for a wide range of values of $\tau$, even for $\tau$ larger than its maximal value reached during the experiment. In the methods based on a grid - search \cite{Lerman1980}, proposed in \cite{Lin2007a}, every point of the curve is treated as a trial contact point. The load - indentation data $\{\delta_i,P_i\}$ are calculated from the corresponding trial in-contact region. Theoretical model of $P(\delta)$ is fitted to $\{\delta_i,P_i\}$ using least squares, while a polynomial is fitted to the corresponding off-contact region. The trail point which gives the lowest total sum of squares of residuals, from both fits, is selected as the final contact point estimate. The trail points located before the actual contact point yield values of $\delta$ and $\tau$ outside their actual range, so that expressions for load valid over wide range of $\tau$ are needed. Formulae for load that combine asymptotic equations and numerical approximations appears to be well suited for this purpose.

Numerical solutions of the problem of a punch indenting a linear - elastic layer have been reported in a few studies, although only for relatively small values of $\tau$. Load required for conical indentation of a layer was calculated numerically in \citep{Jaffar1995} and tabulated for selected values of $\tau \leqslant 20$. These results are in very good agreement with our numerical computations, with discrepancy below $0.053\%$. Load necessary for paraboloidal indentation, plotted for $\tau \leqslant 7$ in \citep{Jaffar1989} also agrees well with our calculations, both for bonded and non-bonded layers. Tabulated values of load required for paraboloidal indentation can be found in
\citep{Hayes1972} for relatively thick  ($\tau \leqslant 3$) bonded layers. The discrepancy between those values and our results is below $0.44\%$. 
\begin{figure}[!t]
	\centering
	\includegraphics{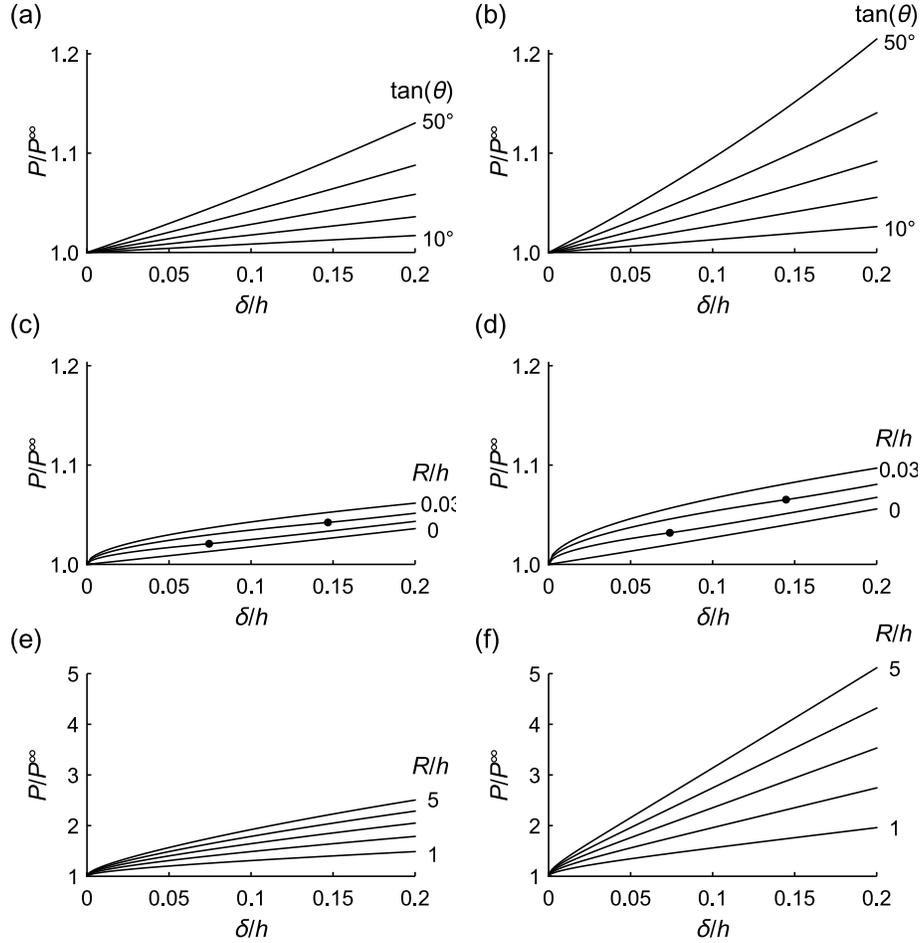}
	\caption{Normalized load $P/P^{\infty}$ required to indent a non-bonded (a, c, e) or bonded (b, d, f) layer, plotted against the ratio of indentation depth $\delta$ and sample's thickness $h$. Load was calculated for conical (a, b), smooth blunt conical  of $20^\circ$ half-angle (c, d) and paraboloidal (e, f) tips. Its values were normalized by load $P^\infty$ calculated under assumption that the sample thickness does not influence the parameters of contact. For a conical tip, $P^\infty$ is given by Sneddon's model \eqref{eq:coneSneddon}, for a blunt conical tip by \eqref{eq:reducedLoadSmoothBluntConeInfiniteThickness}, while for paraboloidal by Hertz's model \eqref{eq:paraboloidHertz}. The black dots in (c, d) mark points on the curves for which $a = b$, i.e. the sample starts to touch the conical body of a blunt tip, so that the tip can no longer be treated as a paraboloid. Note that the vertical axis range in (a) - (d) differs from that in (e) - (f).}
	\label{fig:figReducedLoadVsRelativeIndentationConeParaboloidLooseAndBondedAxisTheSame}
\end{figure}

We also compared several published approximate formulae for load, derived for the problems specified by the boundary conditions \eqref{eq:axiSymContactBoundaryA} - \eqref{eq:axiSymContactBoundaryFiniteBondedB}, with the Nystr\"{o}m method solutions of the corresponding integral equations \eqref{eq:finiteSampleDerZA}. The asymptotic equations for load required to indent a layer with a conical punch, published in \citep{Argatov2011}, deviate less than $10\%$ from the numerical solution provided that $\varkappa < 3.86$ (for a non-bonded) or $\varkappa < 0.68$ (for a bonded later). For tips with half-angle $\theta < 34.4^\circ$, the value of $\varkappa$ is always below $0.68$, even if the tip completely penetrates the layer.  However, similar equations proposed in \citep{Gavara2012} differ significantly from the numerical solutions and the approximate formulae known from the literature, especially in the case of a bonded layer (fig. \ref{fig:plotErrorsOfLiteratureFormulasConeAndParaboloidBonded} a). The equations most commonly used for analysis of force - distance curves recorded with paraboloidal or spherical tips were published in \citep{Dimitriadis2002}. They can be used to correct the substrate effect provided that the layer is relatively thick and the radius of curvature of the tip is small. For an incompressible layer, the relative error of those equations is below $10\%$ when $\varkappa < 0.39$ (non-bonded, fig. \ref{fig:plotErrorsOfLiteratureFormulasConeAndParaboloidLoose} c) or $\varkappa < 0.31$ (bonded layer, fig. \ref{fig:plotErrorsOfLiteratureFormulasConeAndParaboloidBonded} c). The discrepancy between the numerical solution and the equations for load, derived in \citep{Aleksandrov1969}, \citep{Jaffar1989} and \citep{Yang2003} for a very thin, non-bonded layer indented with a paraboloid, quickly decreases with $\varkappa$ (fig. \ref{fig:plotErrorsOfLiteratureFormulasConeAndParaboloidLoose} d). However, an analogous equation derived in \citep{Chadwick2002} and quoted as eq. (15) in \citep{Dimitriadis2002} differs from our numerical results by a factor of $1/2$. The approximate formulae proposed in \citep{Matthewson1981}, \citep{Jaffar1989} and \citep{Yang2003} for a bonded, incompressible layer agree with the numerical results, although the decrease of the relative error is slow (fig. \ref{fig:plotErrorsOfLiteratureFormulasConeAndParaboloidBonded} d).

\begin{figure}[!t]
	\centering
	\includegraphics{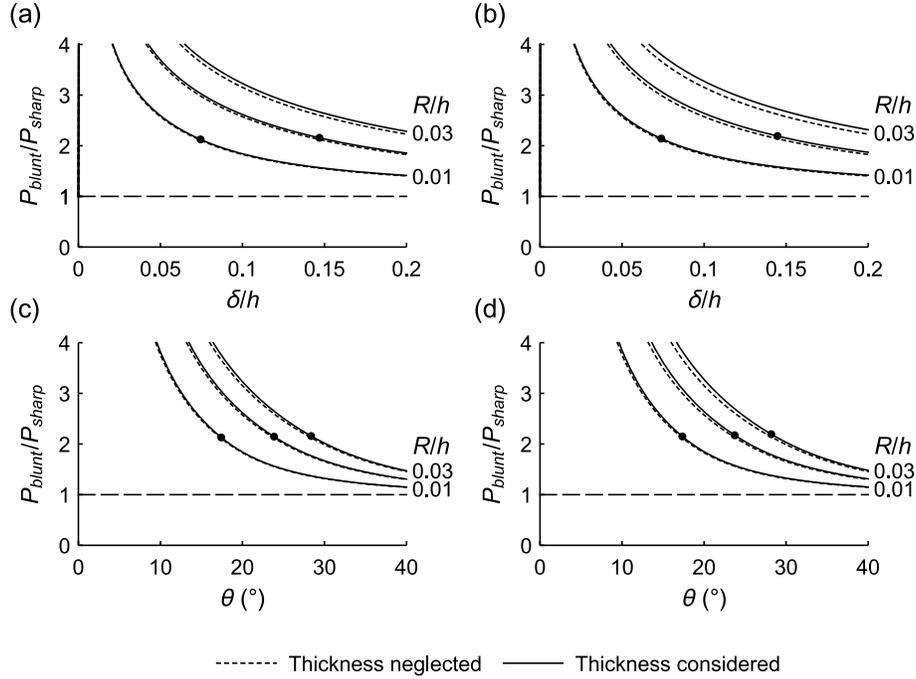}
	\caption{The ratio $P_{blunt}/P_{sharp}$ between values of load required to indent a layer with a smooth blunt conical tip ($P_{blunt}$) and a sharp conical tip ($P_{sharp}$). The calculations were performed for a non-bonded (a, c) and bonded (b, d) layers. (a, b) The relation between $P_{blunt}/P_{sharp}$ and the normalized indentation depth $\delta/h,$ calculated for tips with half angle of $20^\circ$ and the normalized apex radius $R/h$ of 0.01, 0.02 or 0.03. (c, d) The relation between $P_{blunt}/P_{sharp}$ and tip half angle $\theta$, calculated for a fixed indentation depth $\delta/h = 0.1$. The black dots mark points on the curves for which $a = b$, i.e. the sample starts to touch the conical body of a blunt tip, so that the tip can no longer be treated as a paraboloid. The dotted lines show the relations calculated under the assumptions of no influence of sample thickness on load.}
	\label{fig:effectOfRoundApexOfConicalTip}
\end{figure}

Based on the solutions of the Lebedev - Ufliand equation, we can formulate general predictions relevant for analysis of force - distance curves, in particular about potential pitfalls in use of large spherical probes for nearly - incompressible thin samples. The advantages of paraboloidal or spherical AFM tips are well known \citep{Sokolov2013}. Spheres of polysterene or glass can be attached at the free end of a  cantilever \citep{Butt1991,Ducker1992,Mahaffy2000}. Such tips are known as colloidal probes. They induce smaller stresses in the material than sharp pyramidal tips. In addition, colloidal probes are axisymmetric, so analysis of their contact with the sample is relatively simple. However, our numerical results indicate that the effect of the rigid substrate is much more pronounced if spherical probes (fig. \ref{fig:figReducedLoadVsRelativeIndentationConeParaboloidLooseAndBondedAxisTheSame} e, f) are used instead of conical tips - sharp (fig. \ref{fig:figReducedLoadVsRelativeIndentationConeParaboloidLooseAndBondedAxisTheSame} a, b)  or capped by an apex of small radius (fig. \ref{fig:figReducedLoadVsRelativeIndentationConeParaboloidLooseAndBondedAxisTheSame} c, d). It is often assumed that the effect of the substrate can be disregarded if the maximal indentation depth is smaller than $10\%$ of sample thickness. For conical punches, the error due to the substrate effect can be regarded as a function of $\frac{\delta}{h}$ for fixed $\theta$ (fig. \ref{fig:figReducedLoadVsRelativeIndentationConeParaboloidLooseAndBondedAxisTheSame} a, b). For paraboloidal punches, the magnitude of the substrate effect can be treated as a function of $\frac{\delta}{h}$ only when  the ratio $\frac{R}{h}$ is fixed. The influence of substrate may be substantial even if indentations are shallow (fig. \ref{fig:figReducedLoadVsRelativeIndentationConeParaboloidLooseAndBondedAxisTheSame} e, f). Thus, it is advisable to always take into account the presence of the rigid substrate during analysis of force - distance curves recorded with a paraboloidal or spherical probe. 

When radius of a colloidal is large compared to the layer thickness, calculation of load necessary to indent a nearly incompressible,boneded layer (i.e. with $\nu$ close to 0.5) requires accurate values of Poisson's ratio of the sample (fig. \ref{fig:plotReducedLoaBondedImportanceOfPoissonRatio} c, d). For hydrogels (\citep{Urayama1993,Takigawa1996, Chippada2010}) and biological tissues (\citep{Shin1999, Trickey2006,Nijenhuis2014}), $\nu$ is usually between 0.35 - 0.5, often close to 0.5. The accurate value of Poisson's ratio of the sample is rarely available during analysis of force - distance curves. Thus, it appears advisable to avoid probes of radius much larger than sample thickness in experiments with nearly-incompressible materials. Another solution could be including Poisson's ratio as an additional fitting parameter during force curve analysis. It is worthwhile to investigate experimentally whether and when such an approach can be used to extract Poisson's ratio of thin layers from AFM recordings.

The apex of pyramidal or conical tips used in AFM is usually rounded. The predicted load $P_{blunt}$ required to indent a sample with a smooth blunt conical tip is substantially larger then load $P_{sharp}$ required to reach the same indentation depth with a sharp conical tip, both for samples of infinite thickness and for thin layers. Thus, if the presence of round apex of a conical tip is neglected, the calculated values of Young's modulus may be overestimated. The ratio $P_{blunt}/P_{sharp}$ grows with the apex radius of curvature, while it decreases with the half angle of the conical part and indentation depth (fig. \ref{fig:effectOfRoundApexOfConicalTip}). The asymptotic equations \eqref{eq:englandConeParaboloidalApexLoad} -- \eqref{eq:englandConeParaboloidalApexIndentation} can be used to calculate load required to indent a thin layer with a smooth blunt conical tip provided that the radius of apex curvature is small compared to thickness of the layer ($R/h \leqslant 0.3$) and the half-angle of the tip does not exceed $34.5^\circ$. These conditions usually hold in AFM experiments. 

\section*{Acknowledgement}

I would like to cordially thank Dr. Justyna Łabuz (Jagiellonian University) for critical reading of the manuscript. I would also like to express my gratitude to Prof. Halina Gabryś (Jagiellonian University) for help during my work on force microscopy topics.
\appendix

\section{Approximations of the kernel of the Lebedev-Ufliand equation}
\setcounter{figure}{0} 
\label{sect:approximationsLebedevUfliandKernel}
To estimate the effect of the rigid substrate on load and contact radius, it is necessary to numerically evaluate the kernel $K(x,t;\tau)$ for multiple pairs of $x$ and $t$. $K$ can be expressed using $\Omega(y;\tau)$, as in \eqref{eq:finiteSampleDerKernelA}. We will show how to approximate $\Omega$ with a linear combination of the Christov functions. 

$\Omega$ is a non-elementary cosine transform of the weight function $\omega$. Approximating $\omega$ with a linear combination of functions that possess elementary cosine transforms, we will obtain an elementary approximation of $\Omega$. We will start with rescaling $\omega(p;\tau)$
\begin{equation}
\label{eq:smallOmegaTilda}
\widetilde{\omega} (x) =\omega(\tau x;\tau)
\end{equation}
In accordance with \eqref{eq:finiteSampleDerJ3} and \eqref{eq:finiteBondedSampleDerG}, $\tau$ in the argument $\tau x$ of $\omega$ cancels with the parameter $\tau$. Thus, $\widetilde{\omega} (x)$ depends only on $x$. The cosine transform of $\widetilde{\omega}$ will be denoted by $\widetilde{\Omega}$
\begin{equation}
\label{eq:largeOmegaTildaA}
\widetilde{\Omega} (w) = \int_{0}^{\infty}\widetilde{\omega} (x)\cos(x w)\dd{x}
\end{equation}
The transforms $\widetilde{\Omega}$ and $\Omega$ are related by 
\begin{equation}
\label{eq:largeOmegaTildaB}
\widetilde{\Omega} (w)=\frac{\Omega(\frac{w}{\tau};\tau)}{\tau}
\end{equation}
As $\widetilde{\omega}(x)$ decreases exponentially with $x$, we will approximate it with a combination of exponential functions. Investigating the problem of a thin layer with the Green's function method, Li i Dempsey \cite{Li1990} proposed approximating $\widetilde{\omega}(x)$ as  
\begin{equation}
\label{eq:lebedevKernelApproxA}
\widetilde{\omega}(x) \approx \sum_{n=1}^{N}a_n\euler^{-n u x} \qquad u>0
\end{equation}
The coefficients $a_n$ were calculated using the least squares method. The Green's function was expressed as a Hankel transform of an expression containing $\widetilde{\omega}$ as a factor. The approximation \eqref{eq:lebedevKernelApproxA} allowed for calculating the Green's function as a combination of elliptic integrals.  

An approximation of $\Omega$, obtained as a cosine transform of \eqref{eq:lebedevKernelApproxA}, contains only elementary functions
\begin{equation}
\label{eq:lebedevKernelApproxB}
\Omega(y;\tau) \approx \tau \sum_{n=1}^{N}a_n\frac{n u}{n^2 u^2 + \tau^2 y^2}
\end{equation}
The basis of functions used in \eqref{eq:lebedevKernelApproxA} is not orthogonal. To simplify calculation of the expansion coefficients, we will use the orthogonal basis of Laguerre functions $\ell_n(x;u)$, defined as products of normalized Laguerre polynomials $\laguerreL{n}(ux)$ and exponentials
\begin{equation}
\label{eq:lebedevKernelApproxE}
\ell_n(x;u)=\euler^{-\frac{ux}{2}}\laguerreL{n}(ux) \qquad u>0,\quad n=0,1,2,\ldots
\end{equation}
The basis $\left\lbrace  \laguerreL{n}(x) \right\rbrace_{n=0}^\infty$ is orthogonal in $[0,\infty)$. For any non-negative integers $n$ and $m$
\begin{equation}
\label{eq:lebedevKernelApproxF}
\int_{0}^{\infty}\ell_n(x;u) \ell_m(x;u)\dd{x}=\frac{1}{u}\delta_{nm}
\end{equation}
where $\delta_{nm}$ is the Kronecker delta. The expansion of $\widetilde{\omega}(x)$ with respect to the basis $\left\lbrace  \ell_n(x;u) \right\rbrace_{n=0}^\infty$, truncated to the first $N + 1$ terms, can be written as
\begin{equation}
\label{eq:lebedevKernelApproxG}
\widetilde{\omega}(x) \approx u\sum_{n=0}^{N}\int_{0}^{\infty}\widetilde{\omega}(s)\ell_n(s;u)\dd{s}\ell_n(x;u) \qquad u>0
\end{equation}
Substituting $\omega$ for $\widetilde{\omega}$ on the left hand side of \eqref{eq:lebedevKernelApproxG}, we obtain
\begin{equation}
\label{eq:lebedevKernelApproxG2}
\omega(p;\tau) \approx u\sum_{n=0}^{N}\int_{0}^{\infty}\widetilde{\omega}(s)\ell_n(s;u) \dd{s}\ell_n\mathopen{}\left(\frac{p}{\tau};u \right)\mathclose{}
\end{equation}
Cosine transforms of $\ell_n$ can be expressed as \citep{Higgins1977}
\begin{gather}
\label{eq:lebedevKernelApproxI}
\int_{0}^{\infty}\ell_n\mathopen{}\left(\frac{p}{\tau};u \right)\mathclose{} \cos(py)\dd{p}=\frac{\tau\frac{u}{2}}{\left(\frac{u}{2} \right)^2 +\left(y\tau \right)^2 }\chebyshevU{2n}\mathopen{}\left(\frac{y\tau}{\sqrt{\left(\frac{u}{2} \right)^2 +\left(y\tau \right)^2 }} \right)\mathclose{}=\\=\frac{2\tau}{u}\christovCC{2n}\mathopen{}\left(y\tau;\frac{u}{2} \right)\mathclose{} \label{eq:lebedevKernelApproxJ}
\end{gather}
where $U_{2n}$ is a Chebyshev polynomial of the second kind of the $2n$-th degree. In Chebyshev polynomials of even degree all odd coefficients are equal to zero. Thus, the expression on the right hand side of \eqref{eq:lebedevKernelApproxI} is a rational function.  $\christovCC{2n}$ denotes an even Christov function \citep{Christov1982,Boyd1990}. It can be defined using Chebyshev polynomials of the second kind 
\begin{equation}
\label{eq:mathematicalRelationsChristovC}
\christovCC{2n}(x;L)=\frac{L^2}{L^2+x^2}\chebyshevU{2n}\left(\frac{x}{\sqrt{L^2+x^2}} \right) 
\end{equation}
where $n$ is a non-negative integer and $L$ is a positive real number. Combining \eqref{eq:finiteSampleDerOmega}, \eqref{eq:lebedevKernelApproxG2} and \eqref{eq:lebedevKernelApproxJ}, we obtain
\begin{equation}
\label{eq:lebedevKernelApproxK}
\Omega(y;\tau) \approx 2\tau\sum_{n=0}^{N}\left[\int_{0}^{\infty}\widetilde{\omega}(s)\ell_n(s;u) \dd{s}\christovCC{2n}\mathopen{}\left(y\tau;\frac{u}{2} \right)\mathclose{} \right] 
\end{equation}

It can be shown that the approximation \eqref{eq:lebedevKernelApproxK} is equal to the truncated expansion of $\Omega$ with respect to the basis of the even Christov functions $\left\lbrace \christovCC{2n} \right\rbrace_{n=0}^{\infty} $. This basis is orthogonal in $(-\infty,\infty)$ with weight equal to 1 \citep{Christov1982}. In accordance with \eqref{eq:finiteSampleDerOmega}, $\Omega$ depends on $\tau$ only through $\omega(p;\tau)$, which in turn depends on $\frac{p}{\tau}$. After change of variables, $\Omega$ can be regarded as a function of $w=y\tau$, denoted as $\widetilde{\Omega}$ and given by \eqref{eq:largeOmegaTildaB}. To approximate $\Omega(y)$ in $(-2,2)$ for $\tau > 0$, we have to approximate $\widetilde{\Omega}(w)$ in $(-\infty,\infty)$. As $\widetilde{\Omega}$ is an even function, it can be expanded with respect to $\left\lbrace \christovCC{2n}(w;\frac{u}{2}) \right\rbrace_{n=0}^{\infty} $
\begin{equation}
\label{eq:lebedevKernelApproxM}
\widetilde{\Omega}(w) \approx \sum_{n=0}^{N}\left[\int_{-\infty}^{\infty}\frac{4}{\pi u}\widetilde{\Omega}(t)\christovCC{2n}\mathopen{}\left(t;\frac{u}{2}\right)\mathclose{}\dd{t}\christovCC{2n}\mathopen{}\left(w;\frac{u}{2}\right)\mathclose{}\right] 
\end{equation}
\begin{figure}[!t]
	\centering
	\includegraphics{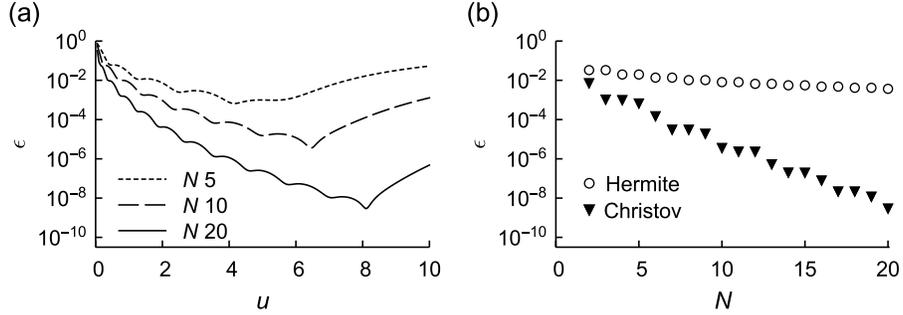}
	\caption{Relative error $\epsilon$ of approximations of $\widetilde{\Omega}$ for a non-bonded layer. (a) The relative error of expansion of $\widetilde{\Omega}$ into $N+1$ even Christov functions, as a function of the scaling factor $u$, for $N$ equal to 5, 10 and 20. (b) The relative error of expansions of $\widetilde{\Omega}$ into series of $N + 1$ even Christov function (triangles) and $N + 1$ even Hermite functions (circles), for $N$ between 2 and 20. The errors were calculated for optimal values of $u$.}
	\label{fig:figureLaguerreHermite}
\end{figure}
We will express \eqref{eq:lebedevKernelApproxM} in a form similar to that of \eqref{eq:lebedevKernelApproxK}. $\widetilde{\Omega}(t)$ and $\christovCC{2n}\left(t;L \right)$ are even, so the integrand in \eqref{eq:lebedevKernelApproxM} is even. The integral in $(-\infty,\infty)$ can be replaced by an integral in $(0,\infty)$. We will also replace $\widetilde{\Omega}$ with $\Omega$, in accordance with \eqref{eq:largeOmegaTildaB}. After simplification
\begin{equation}
\label{eq:lebedevKernelApproxN}
\Omega(y;\tau) \approx 2\tau\sum_{n=0}^{N}\left[\int_{0}^{\infty}\frac{4}{\pi u}\widetilde{\Omega}(t)\christovCC{2n}\mathopen{}\left(t;\frac{u}{2} \right)\mathclose{} \dd{t}\christovCC{2n}\mathopen{}\left(y\tau;\frac{u}{2}\right)\mathclose{}  \right] 
\end{equation}
In accordance with Parseval's theorem for cosine transforms, for any smooth functions $g$ and $h$, which are integrable and square-integrable, it holds that
\begin{equation}
\label{eq:parsevalConsineTransform}
\int_{0}^{\infty}g(s)h(s)\dd{s}=\frac{2}{\pi}\int_{0}^{\infty}\left(\int_{0}^{\infty}g(s)\cos(st)\dd{s}\int_{0}^{\infty}h(s)\cos(st)\dd{s} \right) \dd{t}
\end{equation}
$\widetilde{\Omega}$ is a cosine transform of $\widetilde{\omega}$, while $\frac{2}{u}\christovCC{2n}\mathopen{}\left(t;\frac{u}{2} \right)\mathclose{}$ is a cosine transform of $\ell_n (s;u)$. Combining \eqref{eq:largeOmegaTildaA} and \eqref{eq:lebedevKernelApproxJ} with Parseval's theorem, we obtain
\begin{equation}
\int_{0}^{\infty}\widetilde{\omega}(s)\ell_n(s;u) \dd{s} =\frac{4}{\pi u}\int_{0}^{\infty}\widetilde{\Omega}(t)\christovCC{2n}\mathopen{}\left(t;\frac{u}{2} \right)\mathclose{}\dd{t}
\end{equation}
Hence the direct expansion \eqref{eq:lebedevKernelApproxN} of $\Omega$ with respect to the basis of the even Christov functions is equal to \eqref{eq:lebedevKernelApproxK}. However, calculating the expansion coefficients from \eqref{eq:lebedevKernelApproxK} is much faster. In \eqref{eq:lebedevKernelApproxK}, the integrand is an elementary function $\widetilde{\omega}$, while in \eqref{eq:lebedevKernelApproxN}, the integrand is itself a non-elementary integral.

The relative error $\epsilon$ of an approximation $\widetilde{\Omega}_N$ was calculated as $\epsilon = \frac{\norm{\widetilde{\Omega}-\widetilde{\Omega}_N}_2}{\norm{\widetilde{\Omega}}_2}$
where $\norm{\cdot}_2$ is the $L^2$ norm.
Numerical results indicate that the relative error of the expansion of $\widetilde{\Omega}$ into the even Christov functions decreases exponentially with the number of terms (Fig. \ref{fig:figureLaguerreHermite}). To illustrate the importance of choice of basis functions for approximation of $\widetilde{\Omega}$, we compared the approximations employing the Christov functions with the expansion in series of the Hermite functions, which are also orthogonal in $(-\infty,\infty)$. The relative error of an expansion into the Hermite functions is larger for any $N$ tested.

\pagebreak

\section{Supporting figures}
\renewcommand{\thefigure}{B\arabic{figure}}
\setcounter{figure}{0} 
\begin{figure}[!h]
	\centering
	\includegraphics{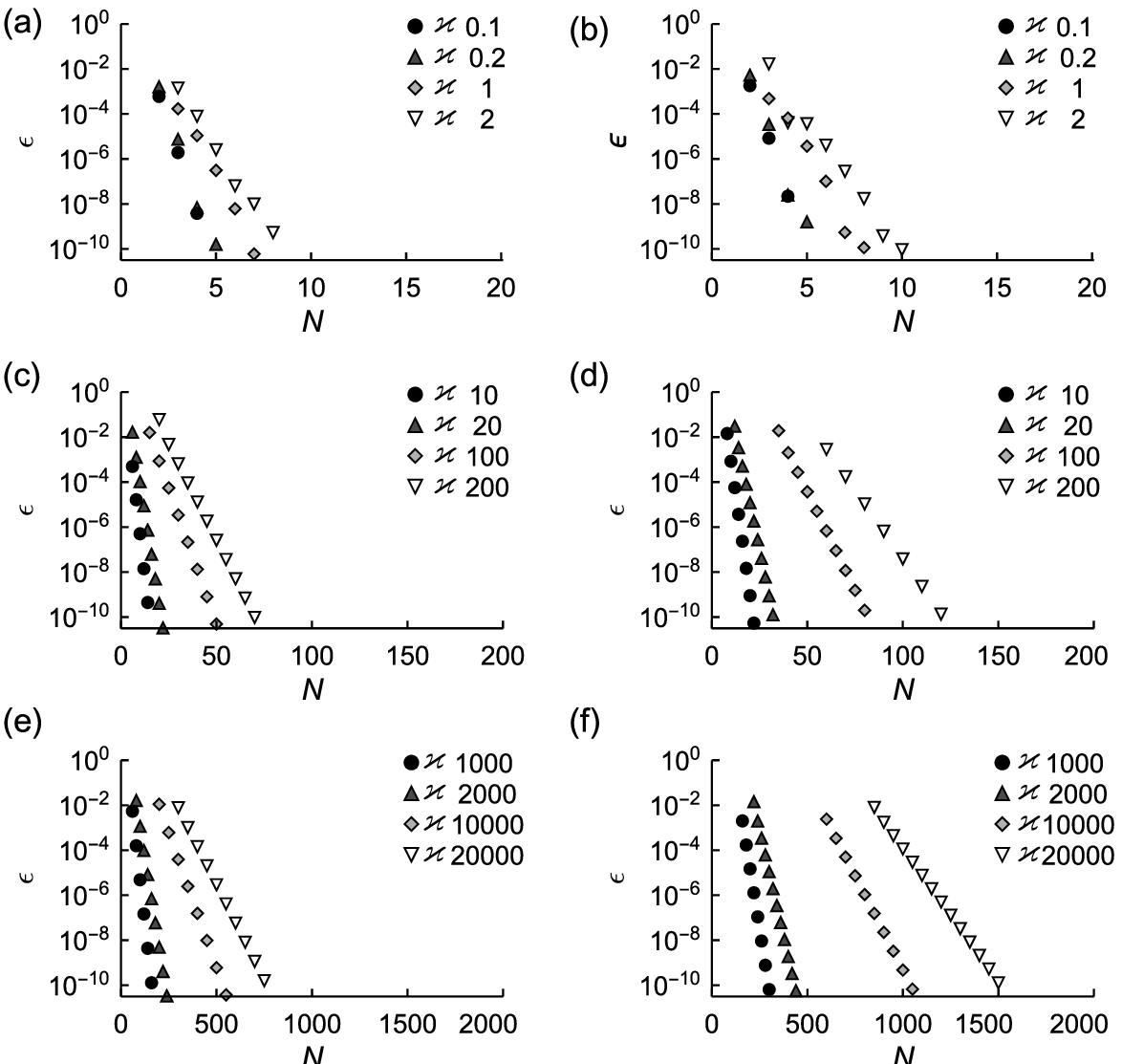}
	\caption{The influence of the number of nodes $N$ in the Gauss - Chebyshev - Radau quadrature on the equilibrium value of $\tau$, calculated from the  Nystr\"{o}m method solutions $\chi$ of the Lebedev - Ufliand equation. The equilibrium value of $\tau$ was found as a root of $\chi(1;\tau)$, using Newton's algorithm. The plotted values are $\epsilon = (\tau_N - \tau_{1600})/\tau_{1600}$, where $\tau_N$ is the estimation obtained with $N$ nodes. Calculations were performed for a  paraboloidal tip indenting a non-bonded (results for different values of reduced indentation depth $\varkappa$ are given in a, c, e) or bonded incompressible (b, d, f) layer.}
	\label{fig:plotErrorContactRadiusVsNLooseAndBondedVariousLambda}
\end{figure}

\begin{figure}[!h]
	\centering
	\includegraphics{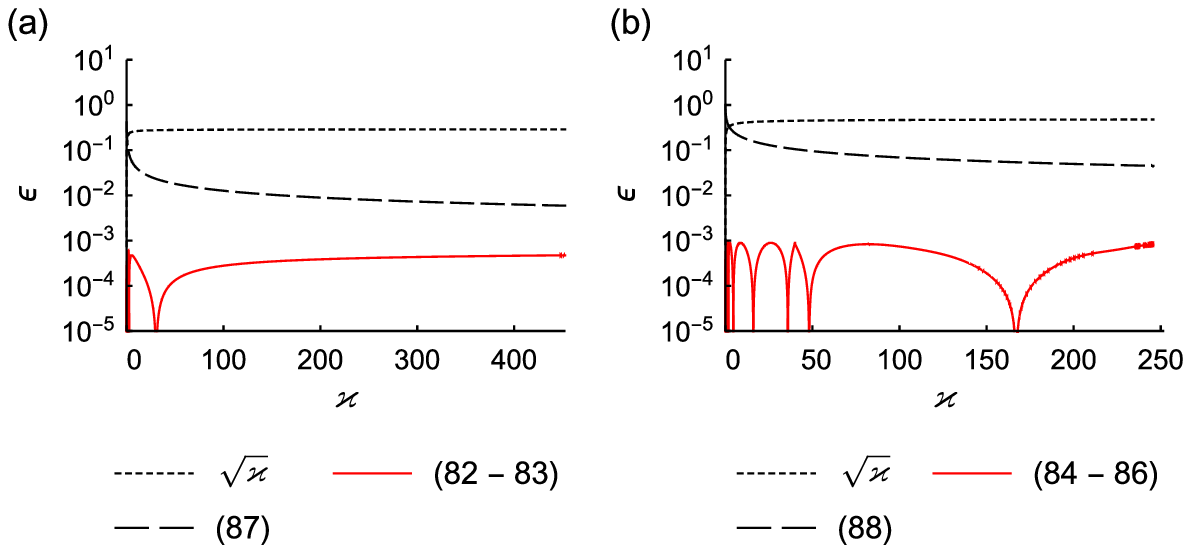}
	\caption{Relative approximation error $\epsilon$ of formulae for dimensionless contact radius $\tau$ for indentation of a non-bonded (a) or bonded incompressible (b) layer with a paraboloidal punch. Red solid lines show the error of the piecewise approximations presented in this work, eq. \eqref{eq:paraboloidTausVsVarKappaA} - \eqref{eq:paraboloidTausVsVarKappaB} in a and \eqref{eq:paraboloidTausVsVarKappaABonded} - \eqref{eq:paraboloidTausVsVarKappaBBonded} in b. Dashed lines show the error of formulae designed for large $\varkappa$, which are \eqref{eq:paraboloidTauLooseVeryThinLayer} in a and \eqref{eq:paraboloidTauBondedVeryThinLayer} in b. Dotted lines show the error made when the equation $\tau = \sqrt{\varkappa}$ is used, which neglects the effect of the rigid substrate.}
	\label{fig:plotsTauErrorsParaboloidLooseAndBoundedIncompressible}
\end{figure}
\pagebreak
  \bibliography{mybibfileNew}

\end{document}